\documentclass[journal]{IEEEtran}
\usepackage{caption}
\usepackage{stfloats}
\usepackage{multirow}
\usepackage{times}
\usepackage{epsfig}
\usepackage{amsmath}
\usepackage{amssymb}
\usepackage[ruled,linesnumbered]{algorithm2e}
\usepackage{graphicx}
\usepackage{algorithmic}
\usepackage{threeparttable}
\usepackage{subfigure}
\usepackage{ifpdf}

\usepackage{booktabs}
\usepackage{threeparttable}
\usepackage{bbding}
\usepackage{pifont}
\newcommand{\cmark}{\ding{51}}%
\newcommand{\xmark}{\ding{55}}%

\usepackage{multicol}
\usepackage{multirow}

\usepackage{color}

\usepackage{makecell} 

\begin{document}
\title{Speech Emotion Recognition via an Attentive Time-Frequency Neural Network}

\author{Cheng~Lu,
        ~Wenming~Zheng$^\dag$,~\IEEEmembership{Senior~Member,~IEEE,}
        ~Hailun~Lian,
        ~Yuan~Zong,~\IEEEmembership{Member,~IEEE,}
        ~Chuangao~Tang,
        ~Sunan~Li,
        and~Yan~Zhao

\thanks{$^\dag$ Corresponding author.}
\thanks{Cheng Lu, Hailun Lian, Sunan Li and Yan Zhao are with the Key Laboratory of Child Development and Learning Science (Ministry of Education), School of Information Science and Engineering, Southeast University, Nanjing 210096, China. (e-mail: cheng.lu@seu.edu.cn).}
\thanks{Wenming Zheng, Yuan Zong and Chuangao Tang are with the Key Laboratory of Child Development and Learning Science (Ministry of Education), School of Biological Science and Medical Engineering, Southeast University, Nanjing 210096, China. (e-mail: {wenming\_zheng, xhzongyuan}@seu.edu.cn).}
}

\markboth{Journal of \LaTeX\ Class Files,~Vol.~, No.~, ~2022}%
{Shell \MakeLowercase{\textit{et al.}}: Bare Demo of IEEEtran.cls for IEEE Journals}

\maketitle

\begin{abstract}

Spectrogram is commonly used as the input feature of deep neural networks to learn the high(er)-level time-frequency pattern of speech signal for speech emotion recognition (SER). \textcolor{black}{Generally, different emotions correspond to specific energy activations both within frequency bands and time frames on spectrogram, which indicates the frequency and time domains are both essential to represent the emotion for SER. However, recent spectrogram-based works mainly focus on modeling the long-term dependency in time domain, leading to these methods encountering the following two issues: (1) neglecting to model the emotion-related correlations within frequency domain during the time-frequency joint learning; (2) ignoring to capture the specific frequency bands associated with emotions.} To cope with the issues, we propose an attentive time-frequency neural network (ATFNN) for SER, including a time-frequency neural network (TFNN) and time-frequency attention. Specifically, aiming at the first issue, we design a TFNN with a frequency-domain encoder (F-Encoder) based on the Transformer encoder and a time-domain encoder (T-Encoder) based on the Bidirectional Long Short-Term Memory (Bi-LSTM). The F-Encoder and T-Encoder model the correlations within frequency bands and time frames, respectively, and they are embedded into a time-frequency joint learning strategy to obtain the time-frequency patterns for speech emotions. Moreover, to handle the second issue, we also adopt time-frequency attention with a frequency-attention network (F-Attention) and a time-attention network (T-Attention) to focus on the emotion-related frequency band ranges and time frame ranges, which can enhance the discriminability of speech emotion features. Extensive experimental results on three public emotional databases, i.\,e., IEMOCAP, ABC, and CASIA, show that the proposed ATFNN outperforms the state-of-the-art methods.

\end{abstract}


\IEEEpeerreviewmaketitle

\section{Introduction}

\IEEEPARstart{S}{peech} emotion recognition (SER) task is to make the machine automatically recognize the emotional state from the human speech. Recently, it has been a research hotspot in affective computing and pattern recognition \cite{schuller2013computational}, \cite{busso2013toward}, \cite{gat2022speaker}, \cite{guo2022learning}, \cite{li2021contrastive}. Generally speaking, the key to deal with the SER problem is to extract the discriminative and generalized features of emotional speech \cite{schuller2009acoustic}, \cite{stuhlsatz2011deep}, \cite{mao2014learning}, \cite{zhang2017speech}, \cite{lu2022domain}. These features mainly contain two categories at present: the hand-crafted features and the deep features.

The hand-crafted features are primarily adopted in the earlier SER works, known as the low-level descriptors (LLDs) \cite{eyben2009openear}, \cite{eyben2010opensmile}, e.\,g., Frame Energy/Loudness, Fundamental Frequency, Zero/Mean-Crossing Rate, Mel-Frequency-Cepstral Coefficients (MFCC), etc. A mount of related works explored various LLDs or these combinations to improve the performance of the SER model. For instance, Schuller et al. \cite{eyben2009openear} integrated 6\,552 hand-crafted features into an open-source toolkit, i.\,e., openEAR, for four affective recognition tasks and then extended the openEAR to a multi-functional and convenient toolkit for diverse speech-related tasks, i\,.e., openSMILE \cite{eyben2010opensmile}. In the earlier works, these LLDs and their combinations were widely utilized as the benchmark features for SER \cite{schuller2009acoustic}, \cite{stuhlsatz2011deep}.

Recently, with the emergence of deep learning, handcrafted features are gradually replaced by deep features extracted by Deep Neural Networks (DNNs), e.\,g. Deep Convolutional Neural Networks (DCNNs) \cite{mao2014learning}, \cite{zhang2017speech}, Recurrent Neural Networks (RNNs) \cite{wang2020speech}, \cite{akccay2020speech}, \cite{xie2019speech}. Instead of LLDs,  these deep features are high(er)-level features. In the extraction of deep features, generally, spectrogram features (e.\,g., MFCC, magnitude spectrogram, and Mel-spectrogram) with rich time-frequency information, are often used as input features of DNNs to learn the high(er)-level time-frequency patterns of speech signal for SER \cite{stuhlsatz2011deep}, \cite{mao2014learning}, \cite{zhang2017speech}, \cite{wang2020speech}. Mao et al. \cite{mao2014learning} investigated to utilize the dimensionality-reduced magnitude spectrogram to learn the salient features by CNNs. Zhang et al. \cite{zhang2017speech} extended the spectrogram to a 3-D Mel-Spectrogram to learn utterance-level speech emotion features from segment features by a discriminant temporal pyramid matching strategy. Wang et al. \cite{wang2020speech} combined the MFCC features and Mel-spectrogram to input a special Long Short-Term Memory (LSTM), called Dual-Sequence LSTM, to extract the utterance-level emotion features.

Furthermore, a time-frequency joint learning strategy is increasingly integrated into DNNs to extract more effective time-frequency representations of speech \cite{sainath2015convolutional}, \cite{li2015lstm}, \cite{satt2017efficient}, \cite{chen20183}. Satt et al. \cite{satt2017efficient} proposed a combined convolution-LSTM network to extract the utterance-level feature with segment-level features. Chen et al. \cite{chen20183} also adopt the 3-D attention-based convolutional recurrent neural network (3-D ACRNN) to represent the utterance level affective-salient features of emotional speech.

Nevertheless, the related works of the time-frequency feature extraction using spectrograms commonly encounter two issues at present. The first issue is that the existing works mainly focus on the correlations between speech frames to characterize the long-term dependency in time domain \cite{mao2014learning}, \cite{wang2020speech}, \cite{li2015lstm}. \textcolor{black}{There are also some works, however, which have revealed that the energy of the frequency band is changing in different emotions \cite{wu2011automatic}, \cite{cowie2001emotion}, \cite{kaiser1962communication}}. In \cite{cowie2001emotion}, Cowie et al. reported the variations of acoustic parameters associated with emotions, and the results of the acoustic spectral parameters indicated that each emotion usually corresponds to some energy activations of specific frequency band ranges on the spectrogram. For instance, the investigation in \cite{cowie2001emotion} reveals that happiness has increased in high-frequency energy, while sadness has decreased in high-frequency energy. These situations indicate that capturing the energy variations in different frequency ranges is also essential for the representation of speech emotions. Therefore, this paper will jointly model the correlations both within the frequency and time domains to obtain robust time-frequency representations of emotions.

\begin{figure*}[t]
\centering
\includegraphics[width=6.5in, height=3.1in]{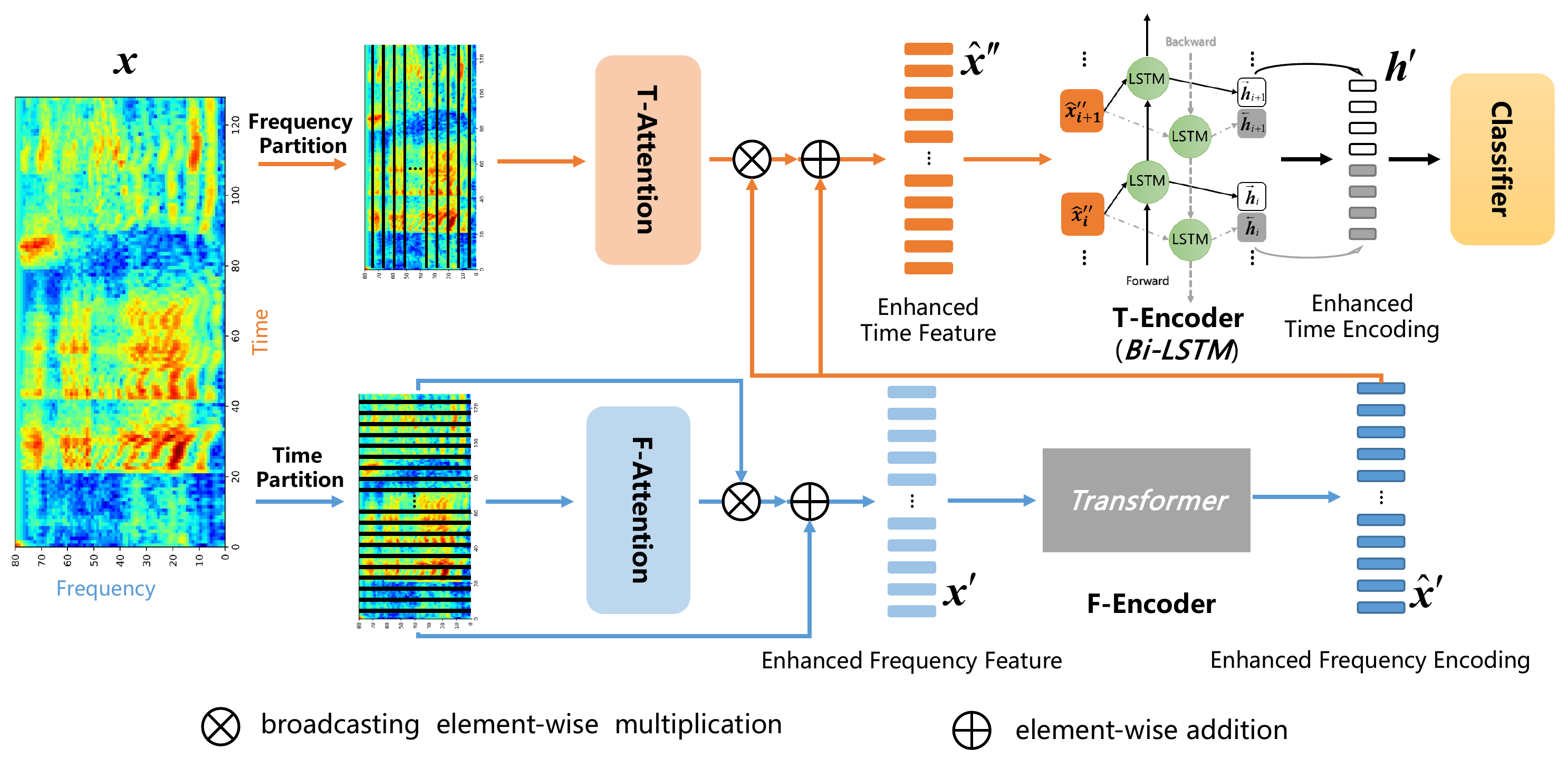}
\caption{Overview of the Attentive Tim-Frequency Neural Network (ATFNN) framework for SER, including a time-frequency neural network (TFNN) and time-frequency attention. The TFNN contains a frequency domain encoder (F-Encoder) and time domain encoder (T-Encoder). The time-frequency attention includes frequency attention (F-Attention) and time attention (F-Attention). In ATFNN, the log-Mel-spectrogram $\boldsymbol{x}$ is firstly encoded as the enhanced frequency feature $\boldsymbol{x}'$ and the enhanced frequency encoding $\hat{\boldsymbol{x}}'$ by F-Attention and F-Encoder (i.\,e., the encoder of Transformer), respectively. Then $\hat{\boldsymbol{x}}'$ is transformed to the enhanced time feature $\hat{\boldsymbol{x}}''$ by T-Attention. Finally, $\hat{\boldsymbol{x}}''$ is as the input of T-Encoder (i.\,e., Bi-LSTM) to generate the enhanced time encoding $\boldsymbol{h}'$ (i.\,e., time-frequency encdoing) for the emotion classification.}
\label{fig:framework}
\end{figure*}

Another issue faced by the time-frequency representation of speech emotion is that the distribution of emotion information in an utterance is sparse in both frequency domain and time domain \cite{chen20183}, \cite{zhang2018attention}. Specifically, it is intuitive that there are some non-speech frames without contents in a long utterance. Thus, not all frames in a sentence contain emotional information. In other words, the frames of an utterance have different contribution degrees in the representation of emotional speech, i.\,e., there are key frame regions highly associated with emotions. For the purpose of capturing these emotion-related regions, Chen et al. \cite{chen20183} investigated adding an attention layer after CNN and LSTM to capture specific temporal frames for the speech emotion representation. Also, Zhang et al. \cite{zhang2018attention} adopted an attention mechanism to empower the sub-layer to focus on emotion salient regions of the spectrogram.

However, these studies mainly embed attention into high-level time-frequency features, leading to introduce the redundant information in the feature extraction and roughly locate the emotion-related regions in the time-frequency domain. In addition, they mainly focus on the selection on the time frames while ignoring the selection on the frequency bands. Actually, some studies have proven that each emotion does not have an energy activation on all frequency bands in the frequency domain \cite{cowie2001emotion}, \cite{kaiser1962communication}, \cite{wu2011automatic}. Therefore, we need to capture the key time frames in the time domain, and the key frequency bands in the frequency domain, which are all highly related to emotions.


To deal with the issues mentioned above, in this paper, we propose an Attentive Time-Frequency Neural Network (ATFNN) to learn the discriminative speech emotion feature for SER, as shown in Figure.\ref{fig:framework}. In detail, we propose a time-frequency neural network (TFNN), including a frequency-domain encoder (F-Encoder) to model frequency features and a time-domain encoder (T-Encoder) to model time features, to capture the correlations within the frequency domain and within the time domain. Moreover, an attention mechanism with a frequency attention network (F-Attention) and a time attention network (T-Attention) is also embedded into TFNN to focus on the specific frequency bands and time frames highly corresponding to emotions.

\textcolor{black}{In summary, the contributions of the paper mainly include the following three points:
\begin{enumerate}
  \item We propose a TFNN to jointly model the correlations both in the time and frequency domains, capturing more emotion information in the time-frequency representation of the speech signal.
  \item We propose a time-frequency attention mechanism, measuring the contribution degree of frequency bands and time frames associated with different emotions, to capture the critical frequency bands and time frames.
  \item By visualizing the attentions in the frequency and time domains, we demonstrate the correlations both within frequency bands and within time frames under different emotions on three public speech emotion databases, i.\,e., IEMOCAP, ABC, and CASIA.
\end{enumerate}
}

The rest of the paper is organized as follows. The proposed ATFNN method is described in detail in Section \uppercase\expandafter{\romannumeral2}. Section \uppercase\expandafter{\romannumeral3} describes the experimental settings and analyzes the experimental results. In the end, Section \uppercase\expandafter{\romannumeral4} concludes the paper and discusses some future works.

\section{ATFNN for SER}

Basically, the ATFNN, shown in Figure.\ref{fig:framework}, is to embed the time-frequency attention strategy (i.\,e., F-Attention and T-Attention) into the TFNN. Thus, in this section, we firstly introduce the TFNN, then describe the ATFNN model in detail.

\subsection{TFNN}

The TFNN aims to learn the high(er)-level time-frequency representation from the input spectrogram feature for SER, which includes two modules, i.\,e., F-Encoder and T-Encoder.

\subsubsection{F-Encoder}

\begin{figure}[t]
\centering
\includegraphics[width=3.3in, height=0.630in]{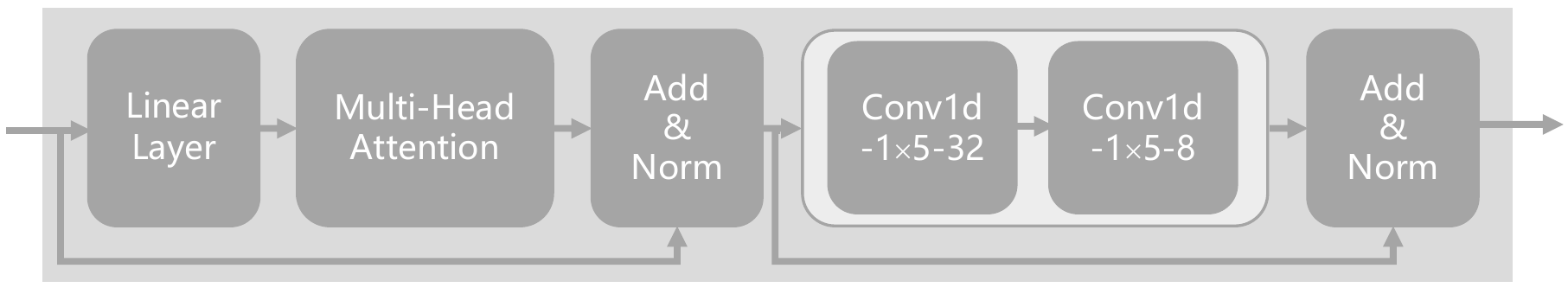}
\caption{Architecture of the F-Encoder, where the input $\boldsymbol{x}$ is firstly transformed to $\boldsymbol{x}^{\prime}$ by a linear layer. Then, after the multi-head self-attention and two convolution layers (with the kernel size is $1\times5$ and the channel number are 32 and 8) and two add-norm layers, the correlations within frequency bands are integrated into the frequency domain encoding $\boldsymbol{\hat{x}}$.}
\label{fig:f-encoder}
\end{figure}

The role of F-Encoder is to characterize the input spectrogram from the frequency domain, including the encoding of frequency information and the correlation modeling within frequency bands. Since different emotions correspond to the activations of specific frequency bands \cite{cowie2001emotion}, the modeling in frequency domain can ignore the order relations between frequency bands. In the Transformer \cite{vaswani2017attention}, the self-attention is adopted to calculate the correlations between tokens in Natural Language Processing (NLP). Inspired by this point, we utilize self-attention to aggregate the correlations across frequency bands into the frequency domain encoding. Thus, we propose an F-Encoder based on the revised encoder of the Transformer to learn the frequency domain representation of emotional speech, shown in Figure. \ref{fig:f-encoder}, in which the position encoding module is abandoned to ignore the order relationship modeling between frequency bands. The proposed F-Encoder is introduced in detail below.

Given the input log-Mel-spectrogram feature of emotional speech as $\boldsymbol{x}\in\mathbb{R}^{f \times t \times 1}$, where $t$ corresponds the frame number and $f$ is the number of frequency bands onto the Mel-scale, the F-Encoder $F_e(\cdot)$ transforms the input $\boldsymbol{x}$ to the frequency domain encoding $\boldsymbol{\hat{x}}\in\mathbb{R}^{f \times t \times c}$, represented as $\boldsymbol{\hat{x}}=F_e(\boldsymbol{x})$. Specifically, the input $\boldsymbol{x}$ is firstly extended the channel information to $\bar{\boldsymbol{x}}\in\mathbb{R}^{f \times t \times c}$ by a linear layer. Then, the multi-head self-attention is utilized to calculate the correlations within frequency bands into frequency features $\bar{\boldsymbol{x}}$. After a series of convolution (with the kernel size is $1\times5$ and the channel number is 8) and add-norm operations \textcolor{blue}{(i.\,e., residual connection operations followed by layer normalization \cite{vaswani2017attention})}, the frequency domain encoding $\boldsymbol{\hat{x}}$ is obtained by integrating the learning of correlations within frequency bands into the time-frequency representation. In addition, it is noting that, $f$, $t$, and $c$ in the paper are set to $80$, $128$, and $8$, respectively.

\subsubsection{T-Encoder}

The T-Encoder needs to capture the long-term dependence in time frames of speech for the representation of emotional information. RNNs, especially LSTM, have demonstrated their powerful capabilities in modeling the long-range dependence of the time series signal, and have been widely used in speech signal processing \cite{wang2020speech}, \cite{akccay2020speech}, \cite{xie2019speech}. Therefore, we adopt a T-Encoder based on a revised Bidirectional LSTM (Bi-LSTM) \cite{wang2020speech}, \cite{xie2019speech}.

The proposed T-Encoder $T_e(\cdot)$ takes the $i^{th}$ frame $\boldsymbol{\hat{x}}_i\in\mathbb{R}^{f \times c}$ of the frequency encoding $\boldsymbol{\hat{x}}$ as the input of each time step in Bi-LSTM to learn the time-frequency representation $\boldsymbol{h}$ of emotional speech. In detail, the time domain encoding process can be defined as $\boldsymbol{h}_i=T_e(\boldsymbol{\hat{x}}_i)$, where $i\in[1,2,...,t]$ represents the frame index, $\boldsymbol{\hat{x}}=\{\boldsymbol{\hat{x}}_i\}_{i=1}^{t}$, and $\boldsymbol{h}_i$ is the $i^{th}$ time-step output of the T-Encoder. Notably, since the T-Encoder is based on three-layer Bi-LSTM with 128 hidden nodes in the paper, the final time-frequency feature $\boldsymbol{h}\in\mathbb{R}^{1 \times 256}$ is obtained by concatenating the last hidden states in the forward direction $\overrightarrow{\boldsymbol{h}_t}\in\mathbb{R}^{1 \times 128}$ and the last hidden states in reversed direction $\overleftarrow{\boldsymbol{h}_t}\in\mathbb{R}^{1 \times 128}$, which can be represented as $\boldsymbol{h}=[\overrightarrow{\boldsymbol{h}_t}, \overleftarrow{\boldsymbol{h}_t}]$.

\subsubsection{TFNN}

The TFNN integrates F-Encoder and T-Encoder into time-frequency joint learning strategy to ensure that the emotion patterns of speech are captured in both frequency domain and time domain. Specifically, the input log-Mel-spectrogram feature $\boldsymbol{x}$ is firstly encoded in frequency bands by F-Encoder to obtain frequency domain encoding $\boldsymbol{\hat{x}}$, represented as $\boldsymbol{\hat{x}}=F_e(\boldsymbol{x})$. Then, $\boldsymbol{\hat{x}}$ is used as the input of T-Encoder to capture the long-term dependence between time frames to obtain the time domain encoding $\boldsymbol{h}$ through a time-frequency joint learning strategy, which can be formalized as $\boldsymbol{h}=T_e(F_e(\boldsymbol{x}))$.

\subsection{ATFNN}

In the ATFNN, two attention networks, i.\,e., F-Attention and T-Attention, are employed in the TFNN to focus on the emotion-related regions in the frequency and time domains.

\subsubsection{F-Attention}

\begin{figure*}[t]
\centering
\includegraphics[width=4.9in, height=2.3in]{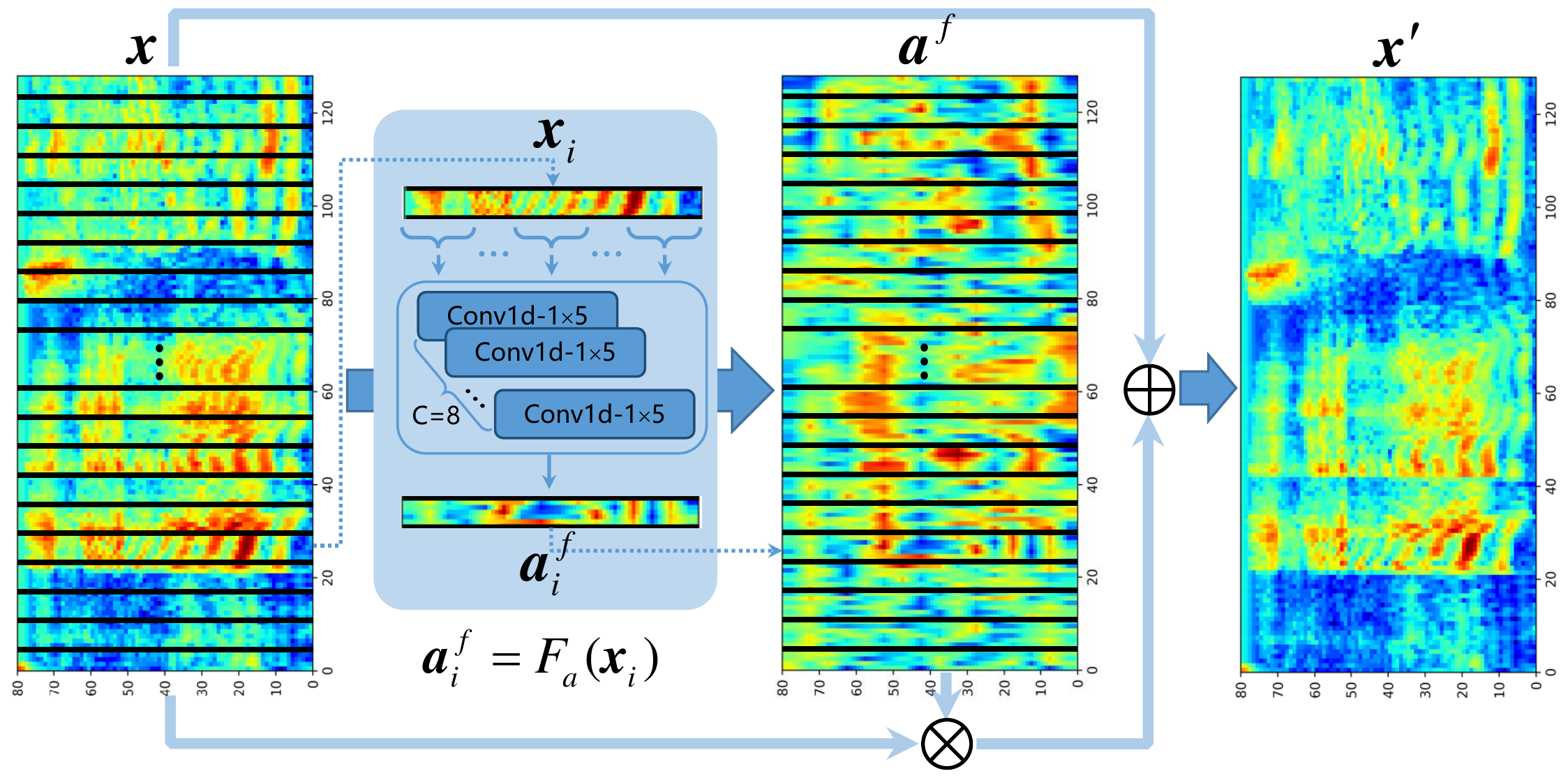}
\caption{Pipeline of the F-Attention in ATFNN, in which the log-Mel-spectrogram $\boldsymbol{x}$ is adopted to calculate the frequency attention weight $\boldsymbol{a}^f_{i}$, then integrate the attention-based frequency feature into $\boldsymbol{x}$ to generate the enhanced frequency feature $\boldsymbol{x}'$. $\otimes$ and $\oplus$ are the same operations in Figure. \ref{fig:framework}.}
\label{fig:f-attention}
\end{figure*}

F-Attention aims to capture salient frequency bands contributing to different emotions, which is implemented by a convolution based block with the kernel size $(1 \times 5 \times c)$, shown in Figure. \ref{fig:f-attention}. To calculate the frequency attention, we firstly represent the input log-Mel-spectrogram $\boldsymbol{x}=\{\boldsymbol{x}_i\}_{i=1}^{t}$ as a set of $\boldsymbol{x}_i\in\mathbb{R}^{f\times c}$, where $i\in[1,2,...,t]$ is the frame index and $\boldsymbol{x}_i$ represents the frequency domain feature of $i^{th}$ frame on $\boldsymbol{x}$. Since the frequency bands related to emotions are within a certain range, e.\,g., high frequency, intermediate frequency, and low frequency \cite{satt2017efficient}, \cite{cowie2001emotion}, we employ 5 frequency bands of $\boldsymbol{x}_i$ as a band group for the convolution to preserve the correlations within frequency bands. Thus, after the convolution operation, the frequency attention weights $\boldsymbol{a}^f_i\in\mathbb{R}^{1 \times (f/5) \times c}$ of different frequency band groups can be obtained, expressed as $\boldsymbol{a}^f_i=F_a(\boldsymbol{x}_i)$. Moreover, as shown in Figure. \ref{fig:f-attention}, we utilize the frequency attention $\boldsymbol{a}^f_i$ to perform a weighted average operation on $\boldsymbol{x}_i$ with respect to the channel $c$ to produce the attention-based frequency features, then integrate these frequency features into the origin input features $\boldsymbol{x}_i$ by the element-wise addition operation to produce the enhanced frequency feature $\boldsymbol{x}_i^{\prime}$, which can be represented as
\begin{equation}
\begin{aligned}
\label{equ:f-attention}
\boldsymbol{x}_i^{\prime}&=\boldsymbol{x}_i \oplus (F_a(\boldsymbol{x}_i) \otimes \boldsymbol{x}_i) \\
                         &=\boldsymbol{x}_i \oplus \left(\frac{1}{c} \sum_{m=1}^{c} {\boldsymbol{a}^f_{i,m} \otimes \boldsymbol{x}_i}\right),
\end{aligned}
\end{equation}
where $\boldsymbol{a}^f_{i,m}\in\mathbb{R}^{1 \times (f/5)}$ represents the attention weight of the $m^{th}$ channel on the frequency attention weight $\boldsymbol{a}^f_i$, and $m\in[1,2,...,c]$. Moreover, $\otimes$ and $\oplus$ represent the broadcasting element-wise multiplication and element-wise addition, respectively. It is noting that, in $\otimes$ of F-Attention, each 5 frequency bands of $\boldsymbol{x}_i$ share the same attention weights for duplication.

After that, we embeds the F-Attention into the F-Encoder, called as the attentive frequency neural network (AFNN), to produce the enhanced frequency encoding $\boldsymbol{\hat{x}}^{\prime}\in\mathbb{R}^{f \times t \times c}$. The AFNN can be defined as \textcolor{red}{$\boldsymbol{\hat{x}}_i^{\prime}=F_e(\boldsymbol{x}_i^{\prime})$,} where $\boldsymbol{\hat{x}}^{\prime}_i\in\mathbb{R}^{f \times c}$ and the enhanced frequency encoding $\boldsymbol{\hat{x}}^{\prime}\in\mathbb{R}^{f \times t \times c}$ is a set of $\boldsymbol{\hat{x}}^{\prime}_i$, written as $\boldsymbol{\hat{x}}^{\prime}=\{\boldsymbol{\hat{x}}^{\prime}_i\}_{i=1}^t$.

\begin{figure*}[t]
\centering
\includegraphics[width=5.6in, height=2.3in]{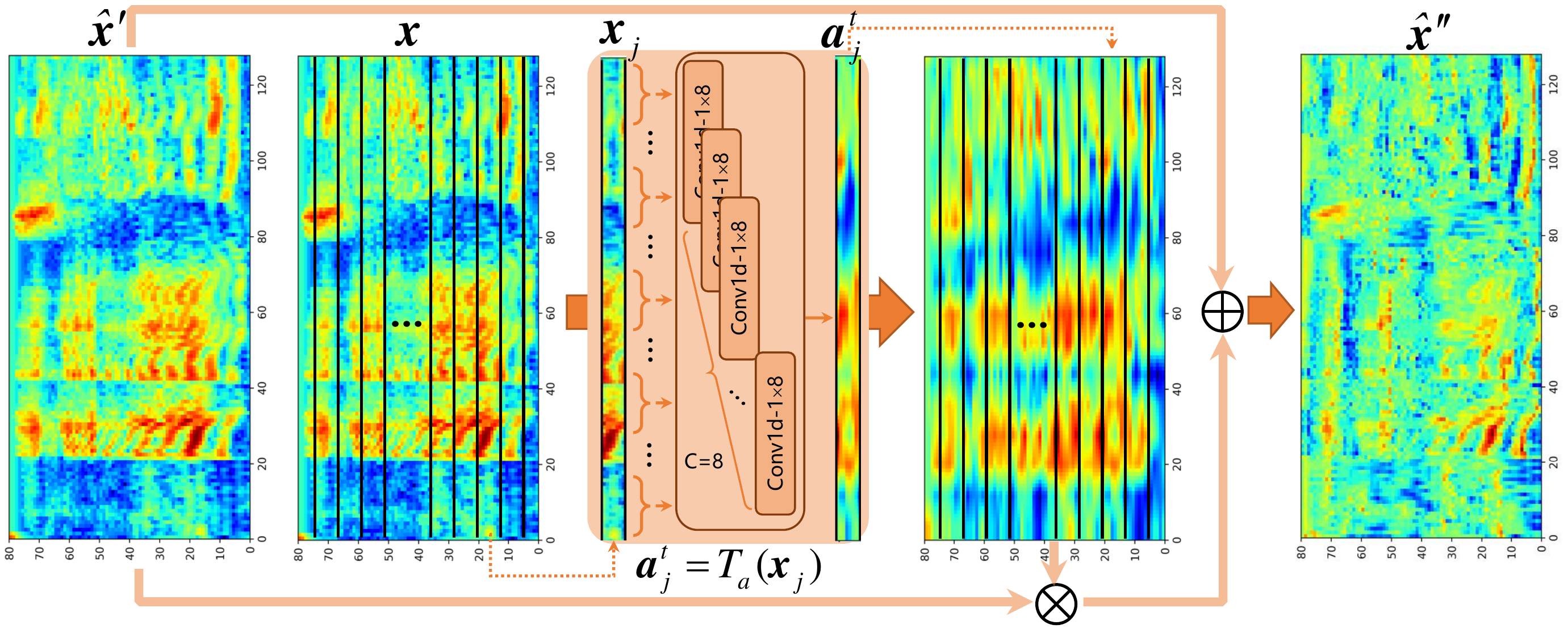}
\caption{Pipeline of the T-Attention in ATFNN, where we also use the log-Mel-spectrogram $\boldsymbol{x}$ to calculate the time attention weight $\boldsymbol{a}^t_{j}$, then integrate the attention-based time feature into the enhanced frequency encoding $\boldsymbol{\hat{x}}^{\prime}$ to generate the enhanced time encoding $\boldsymbol{\hat{x}}^{\prime\prime}$ for the final speech emotion classification. $\otimes$ and $\oplus$ are the same operations in Figure. \ref{fig:framework}.}
\label{fig:t-attention}
\end{figure*}

\subsubsection{T-Attention}

Similar to F-Attention, the T-Attention is also performed by a convolution based block with the kernel size $(1 \times 8 \times c)$ to capture salient frames that contribute to emotions, shown in Figure. \ref{fig:t-attention}. To calculate the time domain attention, we firstly divide the input lo-Mel-spectrogram $\boldsymbol{x}$ into multiple $\boldsymbol{x}_j$ in frequency domain, where $j\in[1,2,...,f]$ represents the $j^{th}$ frequency band of the input $\boldsymbol{x}$ and $\boldsymbol{x}_j\in\mathbb{R}^{t \times c}$ is the time domain feature of the $j^{th}$ frequency band on $\boldsymbol{x}$, i.\,e, $\boldsymbol{x}=\{\boldsymbol{x}_j\}_{j=1}^f$. In T-Attention, 8 time frames of $\boldsymbol{x}_j$ are employed as a frame group for the convolution to obtain the time attention weight $\boldsymbol{a}_j^t\in\mathbb{R}^{1 \times (t/8) \times c}$, expressed as $\boldsymbol{a}_j^t=T_a(\boldsymbol{x}_j)$. Then, as shown in Figure. \ref{fig:t-attention}, we also employ the time attention $\boldsymbol{a}^t_j$ to perform a weighted average operation on $\boldsymbol{x}_j$ with respect to the channel $c$ to generate the attention-based time features, and integrate them into the enhance frequency encoding $\boldsymbol{\hat{x}}_j$ by the element-wise addition operation to obtain the enhanced time feature \textcolor{red}{$\boldsymbol{\hat{x}}_j^{\prime\prime}\in\mathbb{R}^{t \times c}$,} which can be written as
\begin{equation}
\begin{aligned}
\label{equ:t-attention}
\boldsymbol{\hat{x}}^{\prime\prime}_j &=\boldsymbol{\hat{x}}^{\prime}_j \oplus (T_a(\boldsymbol{x}_j) \otimes \boldsymbol{\hat{x}}^{\prime}_j) \\
                                      &=\boldsymbol{\hat{x}}^{\prime}_j \oplus \left(\frac{1}{c} \sum_{n=1}^{c} {\boldsymbol{a}^t_{j,n} \otimes \boldsymbol{\hat{x}}^{\prime}_j}\right),
\end{aligned}
\end{equation}
where $\boldsymbol{a}^t_{j,n}\in\mathbb{R}^{1 \times (t/8)}$ represents the attention weight of the $n^{th}$ channel on the time attention weight $\boldsymbol{a}^t_j$, $n\in[1,2,...,c]$, and the enhanced time feature $\boldsymbol{\hat{x}}^{\prime\prime}\in\mathbb{R}^{f \times t \times c}$ is a set of $\boldsymbol{\hat{x}}^{\prime\prime}_j$, i\,.e., \textcolor{red}{$\boldsymbol{\hat{x}}^{\prime\prime}=\{\boldsymbol{\hat{x}}^{\prime\prime}_j\}_{j=1}^f$.} In $\otimes$ of T-Attention, each 8 time frames of $\boldsymbol{\hat{x}}^{\prime}_j$ share the same attention weights for duplication. Since the T-Encoder is encoded according to the time frame, the ATFNN can also be expressed as $\boldsymbol{h}^{\prime}_i=T_e(\boldsymbol{\hat{x}}^{\prime\prime}_i)$, where $\boldsymbol{h}^{\prime}=\{\boldsymbol{h}^{\prime}_i\}_{i=1}^{t}$ is the enhance time encoding.

\subsubsection{ATFNN}

Integrating F-Attention and T-Attention into F-Encoder and T-Encoder respectively, we extend TFNN to attention-based TFNN, i.\,e., ATFNN. With F-Attention, the ATFNN can fucus on the key frequency bands related to emotions to generate the enhanced frequency domain coding $\boldsymbol{\hat{x}}^{\prime}$, expressed as
\begin{equation}
\label{equ:enhanced-frequency-encoding}
\boldsymbol{\hat{x}}^{\prime}=F_e(\boldsymbol{x} \oplus F_a(\boldsymbol{x})).
\end{equation}
Then, the T-Attention is adopt to capture key time frames related to emotions to obtain a more discriminative time-frequency representation $\boldsymbol{h}^{\prime}_t$, which can be formalized as
\begin{equation}
\begin{aligned}
\label{equ:enhanced-time-encoding}
\boldsymbol{h}^{\prime}_t &= T_e(\boldsymbol{\hat{x}}^{\prime} \oplus T_a(\boldsymbol{\hat{x}}^{\prime})) \\
                          &= T_e(F_e(\boldsymbol{x} \oplus F_a(\boldsymbol{x})) \oplus T_a(F_e(\boldsymbol{x} \oplus F_a(\boldsymbol{x})))).
\end{aligned}
\end{equation}
Finally, the final time-domain encoding $\boldsymbol{h}^{\prime}$ can be regarded as an enhanced time-frequency representation to input into the classifier for SER. Actually, the ATFNN can be regarded as the TFNN that embeds time-frequency attention into the process of time-frequency joint learning.

\section{Experiments}

In this section, we describe the used speech emotion databases, then analyze and discuss the experimental results on our proposed ATFNN.

\subsection{Experimental Databases}

Three public emotional speech databases, i.\,e., the Interactive Emotional Dyadic Motion Capture database (IEMOCAP)  \cite{busso2008iemocap}, the Airplane Behaviour Corpus (ABC) \cite{schuller2007audiovisual}, and China Emotional Database (CASIA) \cite{wang2015speech}, \cite{zhang2008design}, are adopted in our experiments to prove the effectiveness of the proposed ATFNN method.

IEMOCAP is a multimodal database with video, speech, and text scripts, collected by the Speech Analysis and Interpretation Laboratory (SAIL) at the University of Southern California (USC). It is recorded in dyadic sessions where 10 actors (5 females and 5 males) perform improvised or scripted scenarios to elicit several emotional expressions, i.e., \emph{angry}, \emph{happy}, \emph{sad}, \emph{neutral}, \emph{frustrated}, \emph{excited}, \emph{fearful}, \emph{surprised}, \emph{disgusted}, and \emph{others}. In the paper, the \textcolor{red}{improvised} sentences with 4 emotions (i.e., \emph{angry}, \emph{happy}, \emph{sad}, and \emph{neutral}) are all used to perform the experiments according to \textcolor{red}{\cite{satt2017efficient}}, which has 2280 speech samples.

ABC is an audiovisual emotion database with the German collected for the particular target application of public transport surveillance. Also, it has 430 speech samples generated by 8 subjects (4 females and 4 males) in gender balance. They were induced to express one of 6 emotions, i.e., \emph{aggressive}, \emph{cheerful}, \emph{intoxicated}, \emph{nervous, neutral}, and \emph{tired}, by the given scripts.

CASIA is a Chinese speech emotion database released by the Institute of Automation of Chinese Academy of Sciences. It totally consists of 9,600 wave files with 6 emotional states, i.\,e., \emph{angry, fear, happy, neutral, sad}, and \emph{surprise}. Notably, in the paper, we adopt 1200 utterances of ABC released publicly for the experiment section. 4 volunteers with 2 males and 2 females are required to simulate these 6 emotions to produce 300 utterances for each emotion. To describe the selected databases clearly, we also list the detailed parameters of three databases in Table. \ref{tab:description-of-databases}.

\begin{table*}
\caption{Description of the selected speech emotion databases, i.\,e., IEMOCAP, ABC, and CASIA, where 'SR' represents the sampling rate.}
\label{tab:description-of-databases}
\centering
\renewcommand{\arraystretch}{1.5}
\begin{tabular}{|c|c|c|c|c|c|}
\hline
Database&Language&\#Emotions                                                                                                                  &\#Speaker  &\#Utterance&SR(Hz)
\\ \hline \hline
IEMOCAP &English &\emph{angry}(289), \emph{happy}(284), \emph{neutral}(1099), \emph{sad}(608)                                                 &10 (5 males)&2280&\textcolor{red}{16000}
\\ \hline
ABC     & German &\emph{aggressive}(95), \emph{cheerful}(105), \emph{intoxicate}(33), \emph{nervous}(93), \emph{neutral}(79), \emph{tired}(25)&8 (4 males)&430&\textcolor{red}{16000}
\\ \hline
CASIA   & Chinese&\emph{angry}(200), \emph{fear}(200), \emph{happy}(200), \emph{neutral}(200), \emph{sad}(200), \emph{surprise}(200)          &4 (2 males)&1200&\textcolor{red}{16000}
\\ \hline
\end{tabular}
\begin{tablenotes}
\footnotesize
\item \#Emotions denotes the sample distribution of emotions; \#Speaker denotes the number of speakers used in the database; \#Utterance denotes the number of utterance samples used in the database; 'SR' represents the sampling rate.
\end{tablenotes}
\end{table*}

\subsection{Experimental Protocol}
To evaluate the performance of the proposed method, we adopt Leave-One-Speaker-Out (LOSO) cross-validation protocol for three selected databases according to \cite{schuller2009acoustic}, \cite{stuhlsatz2011deep}. Specifically, as shown in Table. \ref{tab:description-of-databases}, the IEMOCAP, ABC, and CASIA databases consist of 10 speakers, 8 speakers, and 4 speakers, respectively. Therefore, in our experiments, the speech samples of one speaker are utilized as the testing data, while the samples of other speakers are used as the training data. It is noting that, since the IEMOCAP includes 5 sessions on 10 speakers, the leave-one-session-out strategy is widely adopt in IEMOCAP database according to \cite{satt2017efficient}, where four sessions (8 speakers) are for training, and one session (2 speakers) is for testing. Therefore, we also select several state-of-the-art methods based on this protocol for comparison \cite{satt2017efficient}, \cite{wu2019speech}, \cite{zhang2018attention}, \cite{bhosale2020deep}, shown in Figure. \ref{tab:si-results}.

Besides, we also utilize two evaluation metrics \cite{schuller2009acoustic}, \cite{stuhlsatz2011deep}, i.\,e., the weighted average recall (WAR) and the unweighted average recall (UAR), to effectively measure the performance of the proposed method, which are commonly adopted in SER task. WAR is known as the 'normal' recognition actuary, while UAR reflects the class-wise recognition accuracy defined as the recall per class divided by the number of classes. Because the selected IEMOCAP and ABC databases are both class-imbalanced in our experiments, the WAR and UAR can better measure the performance of comparison methods.

\subsection{Experimental Setting}
\textcolor{blue}{To perform experiments conveniently, we make the preprocessing for the speech samples. In detail, following the works in \cite{zhang2017speech}, \cite{mao2014learning}, \cite{satt2017efficient}, speech utterances are all divided into small segments with 128 frames (i.\,e., 20ms), which can not only preserve the completeness of speech emotion but also augment the dataset. Then, the log-Mel-spectrogram is extracted by the Short-Time Fourier Transform (STFT), in which 20 ms Hamming window size with 50\% frame overlapping is adopted and 512-point FFT is used on each frame. Besides, the number of Mel-filter bands is set as 80.}

The proposed ATFNN is implemented by the deep learning framework of Pytorch with NVIDIA GeForce RTX3090 GPUs, which is trained from scratch with the batch size of 128 and optimized by the Adam Optimizer with the initialized learning rate of 0.0005.

\subsection{Results and Analysis}

\subsubsection{Results on IEMOCAP}
To compare the proposed ATFNN model with other methods, we select several state-of-the-art works on the IEMOCAP database, i.\,e., DNN-HMM using an alignment generated from the subspace based Gaussian Mixture Model based HMMs (DNN-HMM\_SGMM-Ali.) \cite{mao2019revisiting}, 3 convolution layers and LSTM with 10Hz grid resolution (CNN$+$LSTM Model) \cite{satt2017efficient}, GRU layer upon CNN layers with sequential Capsules (CNN\_GRU-SeqCap) \cite{wu2019speech}, attention based fully convolutional network (FCN$+$Attention Model) \cite{zhang2018attention}, fusion model of acoustic and linguistic features (Model-3\_fusion) \cite{bhosale2020deep}, and an adaptive domain-aware representation learning based model (ADARL) \cite{fan2020adaptive}.

The experimental results on the IEMOCAP database are shown in the Table. \ref{tab:si-results}. From these results, it is obvious that our proposed ATFNN achieves the state-of-the-art performance. Specifically, our ATFNN obtains the best result on WAR (73.81$\%$) than all comparison methods, and it also obtains suboptimal results on UAR (64.48$\%$) than ADARL (65.86$\%$) \cite{fan2020adaptive}. It is worth noting that, since ADARL adds a domain adaptation strategy to eliminate the discrepancy between speakers of the training data and the testing data, it obtains better performance than ATFNN on UAR. Nevertheless, ADARL depends on the hypothesis that the distribution between training and testing data is within a certain upper error bound, such that its universality is worse than our ATFNN.

\begin{figure*}[h]
\centering
\subfigure[Confusion matrix of ATFNN on IEMOCAP]{
\label{iemocap-cm}
\includegraphics[width=0.31\linewidth]{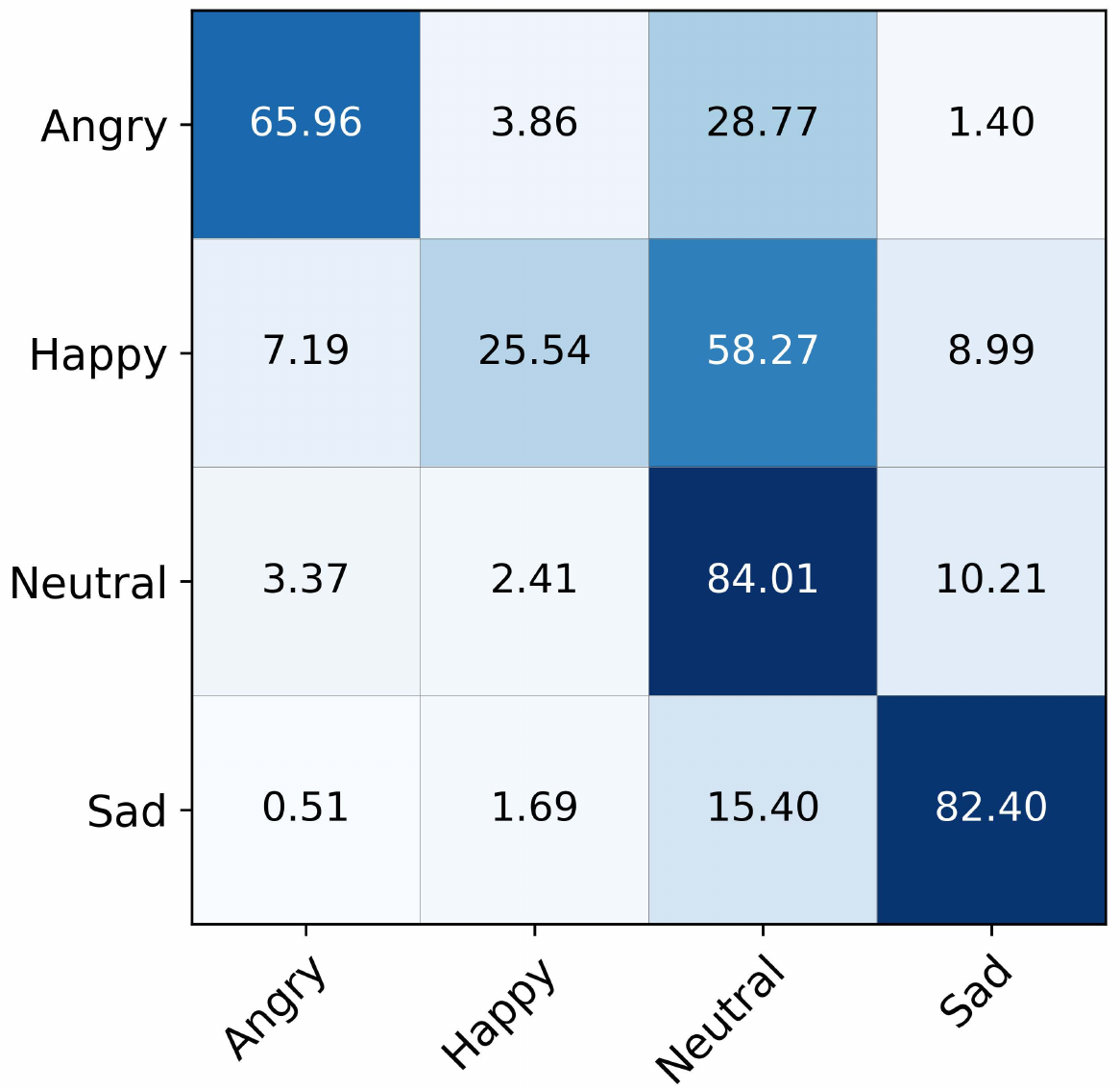}}
\subfigure[Confusion matrix of ATFNN on ABC]{
\label{abc-cm}
\includegraphics[width=0.31\linewidth]{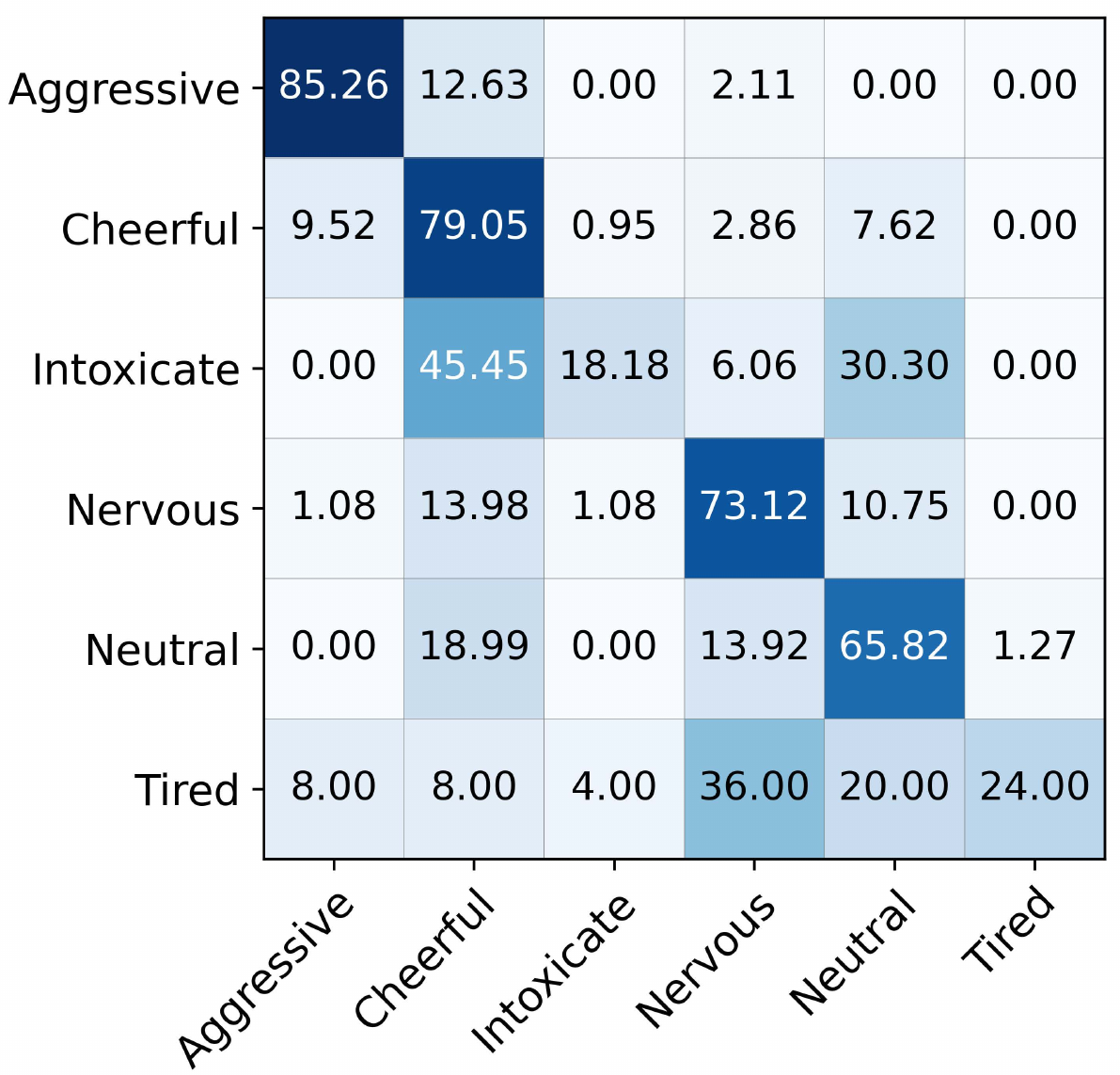}}
\subfigure[Confusion matrix of ATFNN on CASIA]{
\label{casia-cm}
\includegraphics[width=0.31\linewidth]{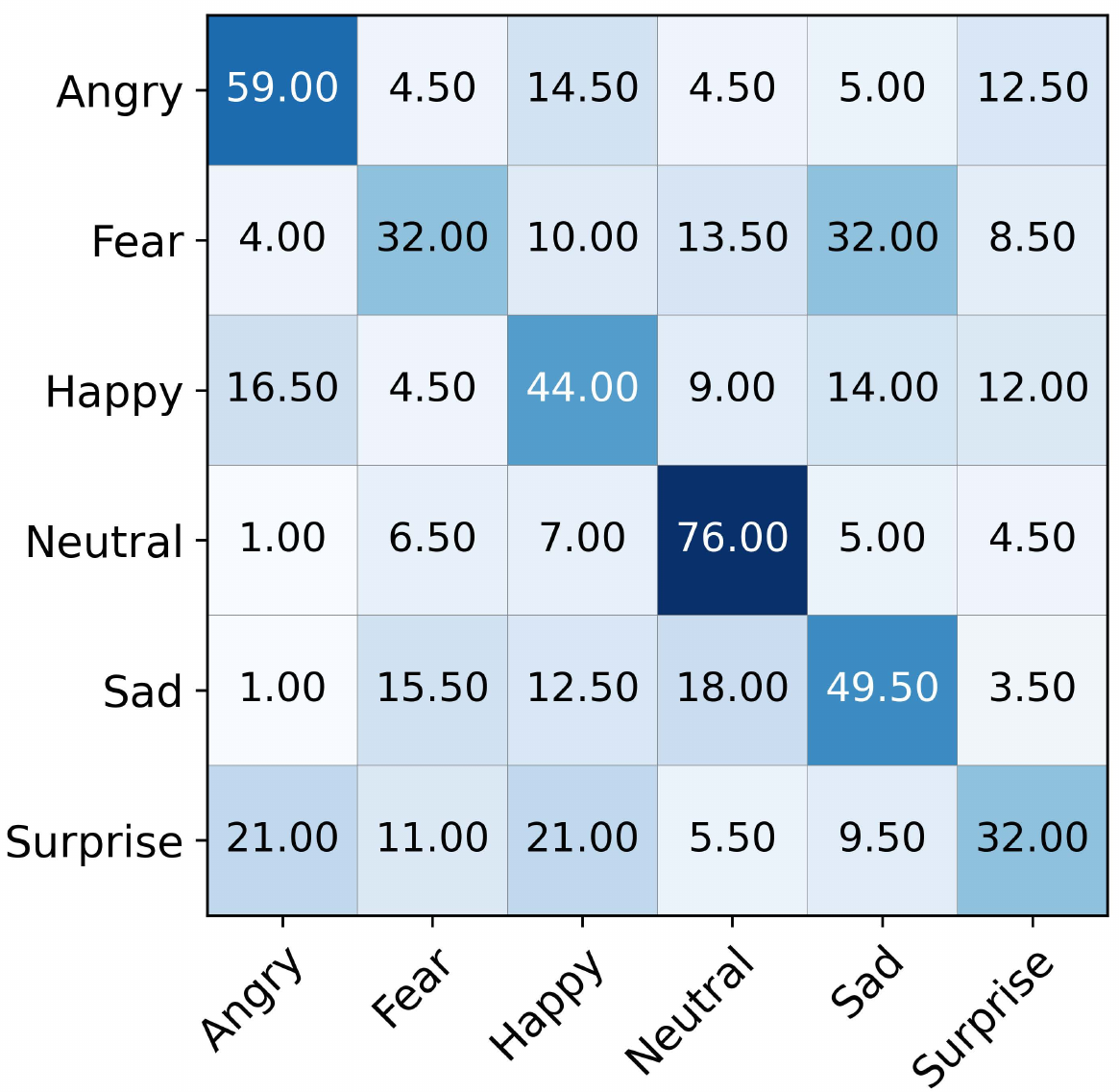}}

\caption{Confusion matrices of ATFNN on three selected speech emotion databases, i.\,e., IEMOCAP, ABC, and CASIA. }
\label{confusion-matrices}
\end{figure*}

\textcolor{black}{Besides, the confusion matrix of ATFNN reported in Figure. \ref{iemocap-cm} also demonstrates the performance of our method in detail, which reveals that ATFNN achieved high recognition accuracies on three emotions, i.\,e., \emph{angry}, \emph{neutral}, and \emph{sad}, while obtains poor performance on \emph{happy}. It is because IEMOCAP is an extremely class imbalance database shown in Table. \ref{tab:description-of-databases}, where \emph{happy} has the smallest number of samples while \emph{neutral} has the largest number of samples. Therefore, this situation may affect the trained classifier to make it easily recognize \emph{happy} as \emph{neutral}.}

\begin{table*}[t]
\caption{Experimental results with WAR and UAR on the three  public speech emotion databases (i.\,e., IEMOCAP, ABC, and CASIA), where the best results are highlight in bold.}
\centering
\renewcommand{\arraystretch}{1.5}
\label{tab:si-results}
\begin{tabular}{|c|c|c|c|c|}
\hline
\multirow{2}{*}{~~Database~~} & \multirow{2}{*}{Experimental Protocol} & \multirow{2}{*}{~~~~Comparison Method~~~~} & \multicolumn{2}{c|}{~~~~~~~~~~Accuracy(\%)~~~~~~~~~} \\ \cline{4-5}
                          &                                            &                                                   & ~~~~~WAR~~~~~   & UAR     \\ \hline \hline
\multirow{7}{*}{IEMOCAP}  & \multirow{7}{*}{\begin{tabular}[c]{@{}c@{}}Leave One Session/Speaker Out (LOSO) \\ (5 Sessions or 10 Speakers)\end{tabular}}
                                                                       & DNN-HMM\_SGMM-Ali. \cite{mao2019revisiting}       & 62.28           & 58.02   \\ \cline{3-5}
                          &                                            & CNN$+$LSTM Model \cite{satt2017efficient}$^*$     & 68.80           & 59.40   \\ \cline{3-5}
                          &                                            & CNN\_GRU-SeqCap \cite{wu2019speech}$^*$              & 72.73           & 59.71   \\ \cline{3-5}
                          &                                            & FCN$+$Attention Model \cite{zhang2018attention}$^*$   & 70.40           & 63.90   \\ \cline{3-5}
                          &                                            & Model-3\_Fusion \cite{bhosale2020deep}$^*$            & 72.34           & 58.31   \\ \cline{3-5}
                          &                                            & ADARL \cite{fan2020adaptive}                      & 73.02           &\textbf{65.86} \\ \cline{3-5} \cline{3-5} \cline{3-5}
                          &                                            & ATFNN (ours)                                      & \textbf{73.81}  & 64.48 \\ \hline\hline

\multirow{5}{*}{ABC}      & \multirow{5}{*}{\begin{tabular}[c]{@{}c@{}}Leave One Speaker Out (LOSO) \\ (8 Speakers)\end{tabular}}
                                                                       & LLDs$+$HMM/GMM \cite{schuller2009acoustic}        & 57.70           & 48.80    \\ \cline{3-5}
                          &                                            & RGSR-LPDA \cite{xu2018connecting}$^{**}$                 & N/A             & 49.40    \\ \cline{3-5}
                          &                                            & LLDs$+$SVM \cite{schuller2009acoustic}            & 61.40           & 55.50    \\ \cline{3-5}
                          &                                            & GerDA \cite{stuhlsatz2011deep}                    & 61.50           & 56.10    \\ \cline{3-5} \cline{3-5} \cline{3-5}
                          &                                            & ATFNN (ours)                                      & \textbf{68.84}  & \textbf{57.57}
                          \\ \hline \hline

\multirow{5}{*}{CASIA}    & \multirow{5}{*}{\begin{tabular}[c]{@{}c@{}}Leave One Speaker Out (LOSO) \\ (4 Speakers)\end{tabular}}
                                                                       & LLD$+$DR \cite{liu2018speech}                     & 39.50           & 39.50     \\ \cline{3-5}
                          &                                            & DNN$+$ELM \cite{han2014speech}                    & 41.17           & 41.17     \\ \cline{3-5}
                          &                                            & HuWSF \cite{sun2015weighted}                      & 43.50           & 43.50     \\ \cline{3-5}
                          &                                            & DTPM \cite{zhang2017speech}                       & 45.42           & 45.42     \\ \cline{3-5} \cline{3-5} \cline{3-5}
                          &                                            & ATFNN (ours)                                      & \textbf{48.75}  & \textbf{48.75} \\ \hline
\end{tabular}
\begin{tablenotes}
\footnotesize
\item $^*$ denotes that the used experiment protocol is leave-one-session-out, i.\,e., four sessions (8 speakers) are for training and one session (2 speakers) is for testing, according to \cite{satt2017efficient};
    $^{**}$ denotes that the samples of 4 speakers are for training and 2 speakers are for testing, according to \cite{xu2018connecting};
    'N/A' indicates the results are not reported in the corresponding paper.
\end{tablenotes}
\end{table*}

\subsubsection{Results on ABC}

For the comparison purpose, we also select several public works on the ABC database, i.\,e., LLDs with HMM/GMM (LLDs$+$HMM/GMM) \cite{schuller2009acoustic}, random anchor points generalized spectral regression with locally penalized discriminant analysis (RGSR-LPDA) \cite{xu2018connecting}, LLDs with SVM (LLDs$+$SVM) \cite{schuller2009acoustic}, and generalized discriminant analysis (GerDA) based on DNNs \cite{stuhlsatz2011deep}.

The results on the ABC database, as shown in Table. \ref{tab:si-results}, also demonstrate that our proposed ATFNN is superior both on WAR (68.84$\%$) and UAR (57.57$\%$) to other comparison methods not only on traditional methods (i.\,e., LLDs$+$HMM/GMM, RGSR-LPDA, and LLDs with SVM), but also on deep learning based methods (i.\,e., GerDA), which proves that the ATFNN can extract more discriminative and instance-adaptive representation of speech emotion on the ABC database. Moreover, we also report the confusion matrix on the ABC database, shown in Figure. \ref{abc-cm}. It is clear that the ATFNN have high recognition accuracies on four emotions, i.\,e., \emph{aggressive}, \emph{cheerful}, \emph{nervous}, and \emph{neutral}, while have poor performance on two emotions, i.\,e., \emph{intoxicate} and \emph{tired}. \textcolor{black}{Specifically, on the ABC, the recognition of \emph{intoxicate} is easily confusing with \emph{cheerful} and \emph{neutral}, whereas \emph{tired} is more recognized as \emph{nervous} and \emph{neutral}. The cause of these results on ABC is similar to reasons for IEMOCAP, that is, the numbers of speech samples on \emph{intoxicate} and \emph{tired} are obviously smaller than other emotions, leading to influence the trained classifier. In addition, another reason may be that the arousal and valance of \emph{intoxicate} and \emph{cheerful} are all close, resulting in the confusing recognition.}

\begin{table*}[t]
\caption{Ablation study of different architectures for ATFNN on the three public speech emotion databases (i.\,e., IEMOCAP, ABC, and CASIA), where the best results are highlight in bold. Moreover, '\cmark' or '\xmark' represents the network with or without the corresponding architecture.}
\centering
\renewcommand{\arraystretch}{1.5}
\begin{tabular}{|c|c|c|c|c|c|c|c|c|c|}
\hline
\multirow{2}{*}{Network}& \multicolumn{3}{c|}{Ablation Architecture}&\multicolumn{2}{c|}{~~~~~IEMOCAP(\%)~~~~~}& \multicolumn{2}{c|}{~~~~~ABC(\%)~~~~~}& \multicolumn{2}{c|}{~~~~~CASIA(\%)~~~~~} \\ \cline{2-10}
      & Attention & F-Encoder & T-Encoder & ~~WAR~~      & ~~UAR~~       & ~~WAR~~      & ~~UAR~~      & ~~WAR~~      & ~~UAR~~       \\ \hline \hline
AFNN  & \cmark    & \cmark    & \xmark    & 70.71        & 59.10         & 58.14        & 48.22        & 32.50        & 32.50         \\ \hline
ATNN  & \cmark    & \xmark    & \cmark    & 71.76        & 61.90         & 66.05        & 55.01        & 46.42        & 46.42         \\ \hline
TFNN  & \xmark    & \cmark    & \cmark    & 72.99        & 63.60         & 67.21        & 55.99        & 46.67        & 46.67         \\ \hline
ATFNN & \cmark    & \cmark    & \cmark    &\textbf{73.81}&\textbf{64.48} &\textbf{68.84}&\textbf{57.57}&\textbf{48.75}&\textbf{48.75} \\ \hline
\end{tabular}
\label{tab:ablation-results}
\end{table*}

\subsubsection{Results on CASIA}

For the fair comparison on the CASIA, four sate-of-the-art methods are chosen in the experiments, including LLDs with a dimension reduction combining PCA and LDA (LLD$+$DR)~\cite{liu2018speech}, DNNs with the extreme learning machine (DNN\_ELM) \cite{han2014speech}, weighted spectral feature learning based on local Hu moment (HuWSF) \cite{sun2015weighted}, and deep convolutional neural network with discriminant temporal pyramid matching (DTPM) \cite{zhang2017speech}.

The experimental results on the CASIA are reported in the Table. \ref{tab:si-results}, which reveal that our proposed ATFNN achieves the highest recognition accuracies than other comparison methods in terms of both WAR (48.75$\%$) and UAR (48.75$\%$). In detail, deep learning based methods, i.\,e., DNN\_ELM and DTPM, outperform the LLDs based traditional method (LLD$+$DR), while our ATFNN is significantly superior than the current sate-of-the-art methods. It is noted that since the number of samples used in each emotion of the CASIA in the paper is equal, the WAR in the experimental results is equal to UAR. Furthermore, the confusion matrix of ATFNN on the CASIA is also calculated to verify the detailed performance of our proposed method, which is given in Figure. \ref{casia-cm}. From these results, we can observe that our ATFNN achieve the high performance on four emotions, i.\,e., \emph{angry}, \emph{happy}, \emph{neutral}, and \emph{sad}, whereas has a certain confusion in the recognition of other two emotions, i.\,e., \emph{fear} and \emph{surprise}. Specifically, \emph{fear} is easier to confuse with \emph{sad}, while \emph{surprise} is always confused with \emph{angry} and \emph{happy}. The reason may be that \emph{fear} and \emph{sad} are relatively close in valence and arousal, causing two emotions to be induced by each other sometimes, while \emph{surprise}, \emph{angry}, and \emph{happy} are the high arousal emotions, which may affect the recognition of each other slightly.

\subsubsection{Experimental Result Discussion}

Throughout all experiments on IEMOCAP, ABC, and CASIA, some results are also worthy of discussing, from which we can investigate the applicability and limitation of our proposed ATFNN. Firstly, two selected databases, i.e., IEMOCAP, ABC, are class-imbalanced, resulting in the discrepancy between WAR and UAR. Moreover, the recognition of different emotions has obvious confusion, as shown in Figure. \ref{confusion-matrices}, where $58.27\%$ of speech samples on \emph{happy} are recognized as \emph{neutral} in IEMOCAP, and $45.45\%$ of samples on \emph{intoxicate} in ABC are recognized as \emph{cheerful}. These issues may be caused by the following reasons: one reason is that these emotions are relatively close in valence or arousal, and the other reason is that the long-tailed distribution of speech samples leads the classifier being biased towards emotions with more samples. Next, although the speech samples of CASIA are class-balanced, its recognition performance is much lower than that of IEMOCAP and ABC. This may be because CASIA is collected by forcing the speaker to express different emotions in the same sentence, limiting the diversity of speech samples. Also, CASIA is based on a tonal language (i.\,e., Chinese) rather than the non-tonal language (i.\,e., English). In addition, since the used LOSO protocol in our experiments is all based on speaker-independent, the results on three databases demonstrate that our ATFNN without any domain adaptation strategy not only preserves the discriminability of emotional speech representation but also eliminates the influence of unrelated information in the emotional speech, e.\,g., speakers, speaking styles.

\subsubsection{Ablation Experiment}

To verify the effectiveness of our proposed framework, the additional experiments are implemented for different architectures of ATFNN, and the results in terms of WAR and UAR are reported in Table. \ref{tab:ablation-results}, where AFNN and TFNN represent the FNN with F-Attention and the TNN with T-Attention respectively, described in Section \uppercase\expandafter{\romannumeral2}. From the ablation results, obviously, we can observe that the attention based time-frequency joint learning method (i.\,e., ATFNN) has \textcolor{red}{better} performance than the time-frequency joint learning method (i.\,e., TFNN), which indicates the proposed attention-based method can focus on the emotion-related parts in the time-frequency domain. In addition, we also observe that the time-frequency based method (i.\,e., ATFNN) also perform better than the time or frequency based method (i.\,e., AFNN or ATNN), which also shows that time-frequency joint learning can indeed obtain the discriminative representation of speech emotion. It is also interesting to see that, although the frequency domain representation is meaningful for emotions, the experimental results also show that ATNN performs better than AFNN.

\subsection{Visualization of F-Attention and T-Attention}

\begin{figure*}[h]
\centering
\subfigure[Spectrogram of \emph{happy} on IEMOCAP]{
\begin{minipage}[b]{0.30\textwidth}
\includegraphics[width=\textwidth]{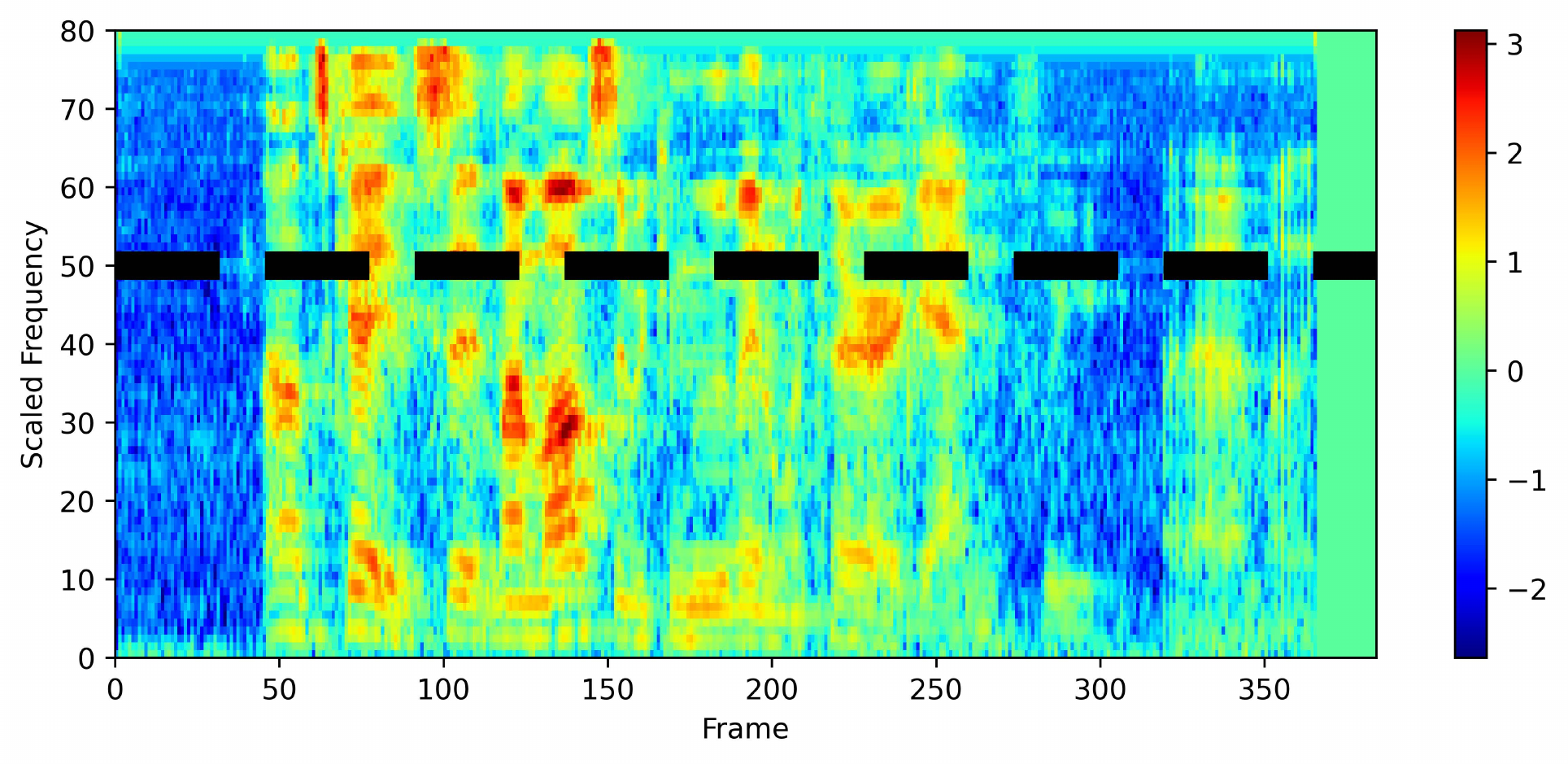}
\end{minipage}
\label{fig:vi-iem-hap-spe}
}
\subfigure[F-Attention of \emph{happy} on IEMOCAP]{
\begin{minipage}[b]{0.30\textwidth}
\includegraphics[width=\textwidth]{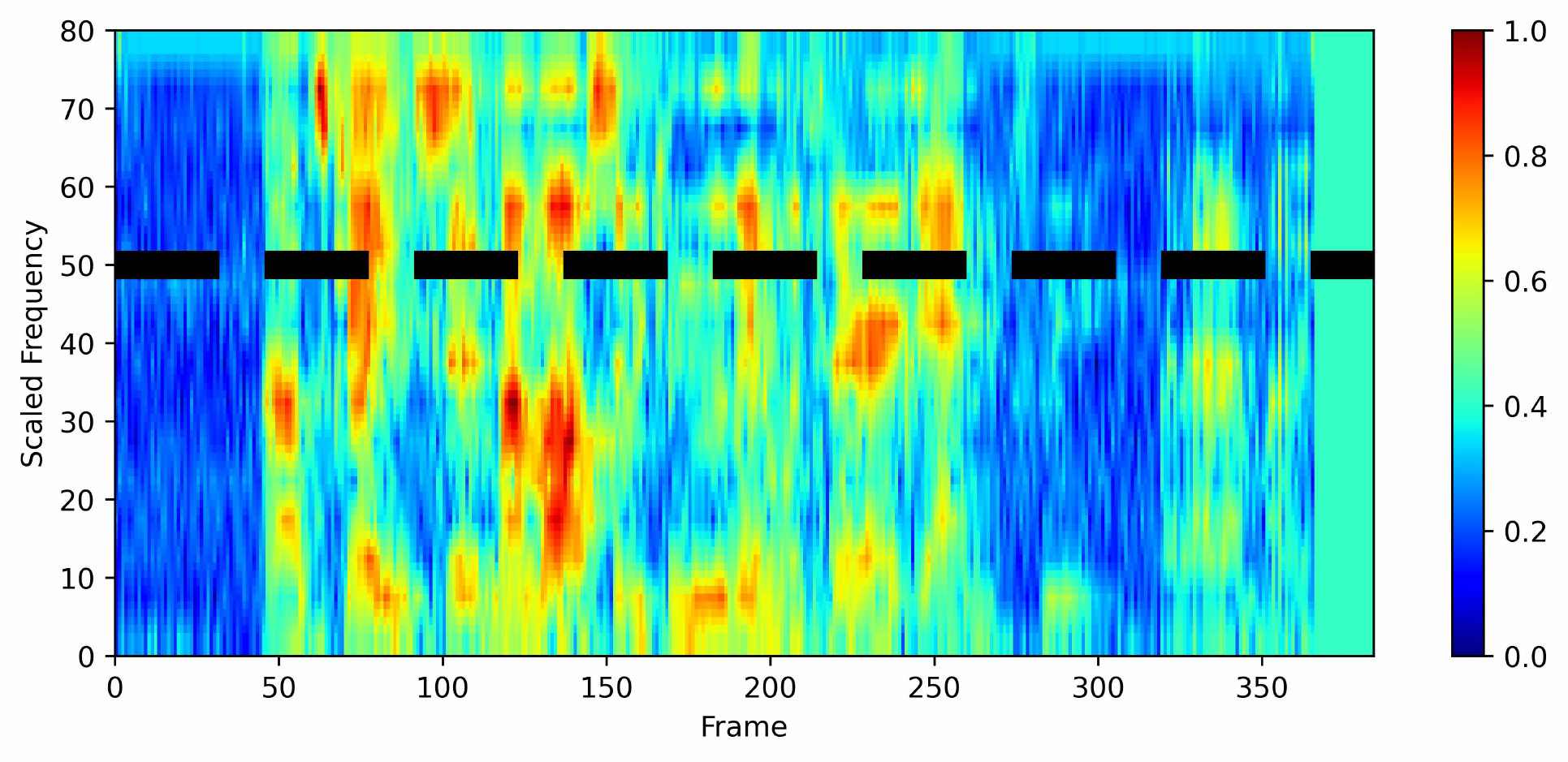}
\end{minipage}
\label{fig:vi-iem-hap-fat}
}
\subfigure[T-Attention of \emph{happy} on IEMOCAP]{
\begin{minipage}[b]{0.30\textwidth}
\includegraphics[width=\textwidth]{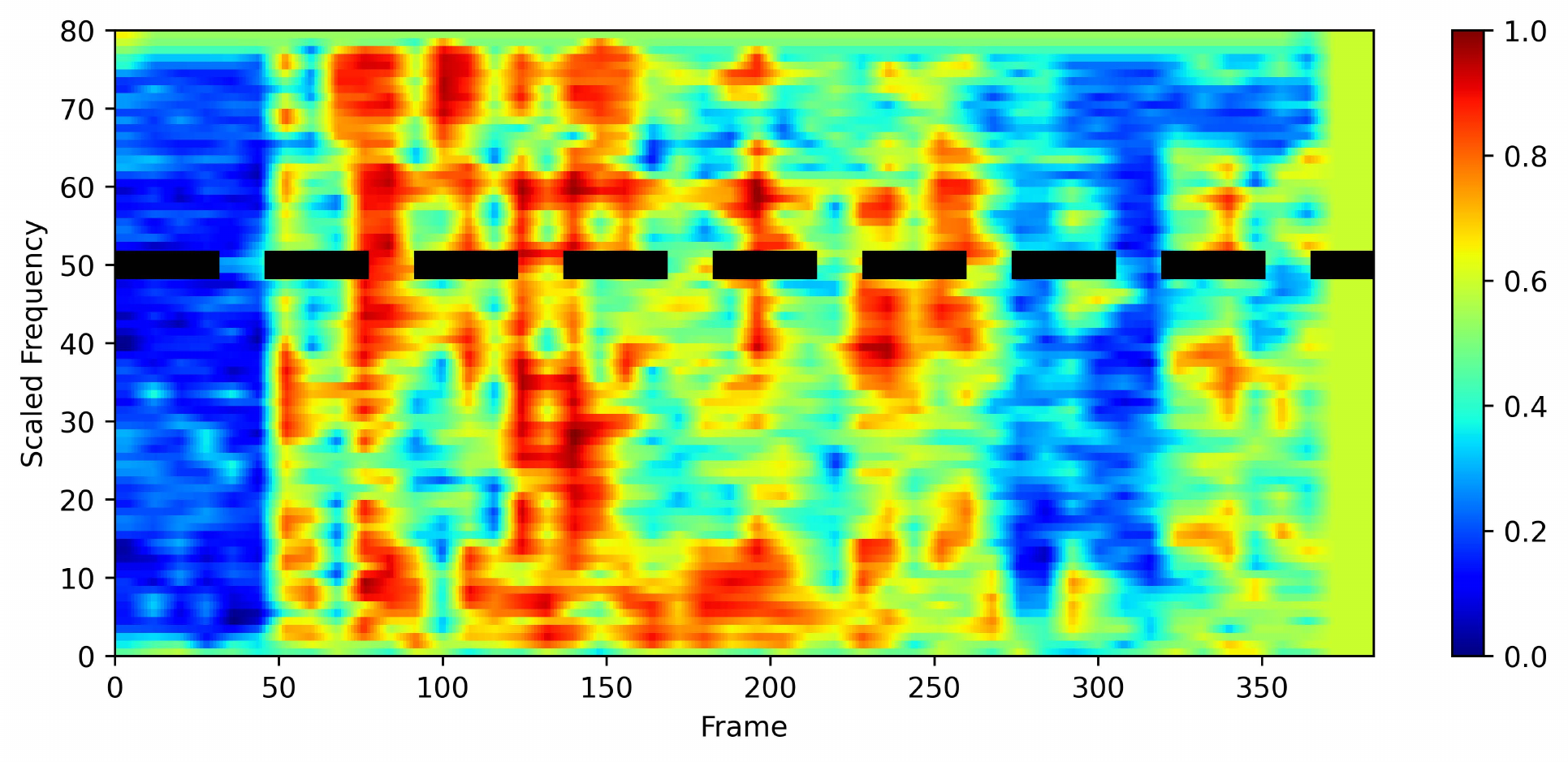}
\end{minipage}
\label{fig:vi-iem-hap-tat}
}

\subfigure[Spectrogram of \emph{sad} on IEMOCAP]{
\begin{minipage}[b]{0.30\textwidth}
\includegraphics[width=\textwidth]{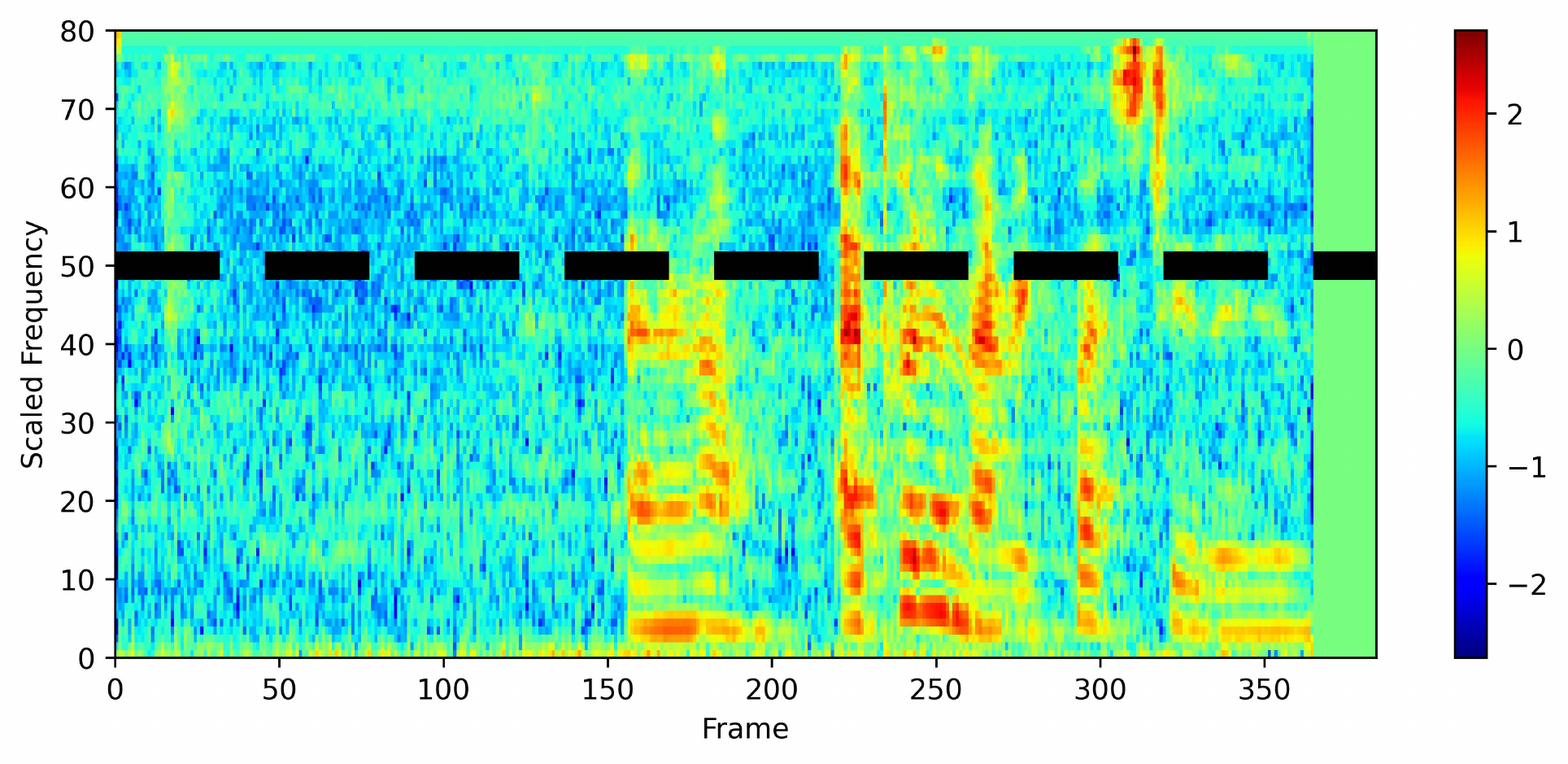}
\end{minipage}
\label{fig:vi-iem-sad-spe}
}
\subfigure[F-Attention of \emph{sad} on IEMOCAP]{
\begin{minipage}[b]{0.30\textwidth}
\includegraphics[width=\textwidth]{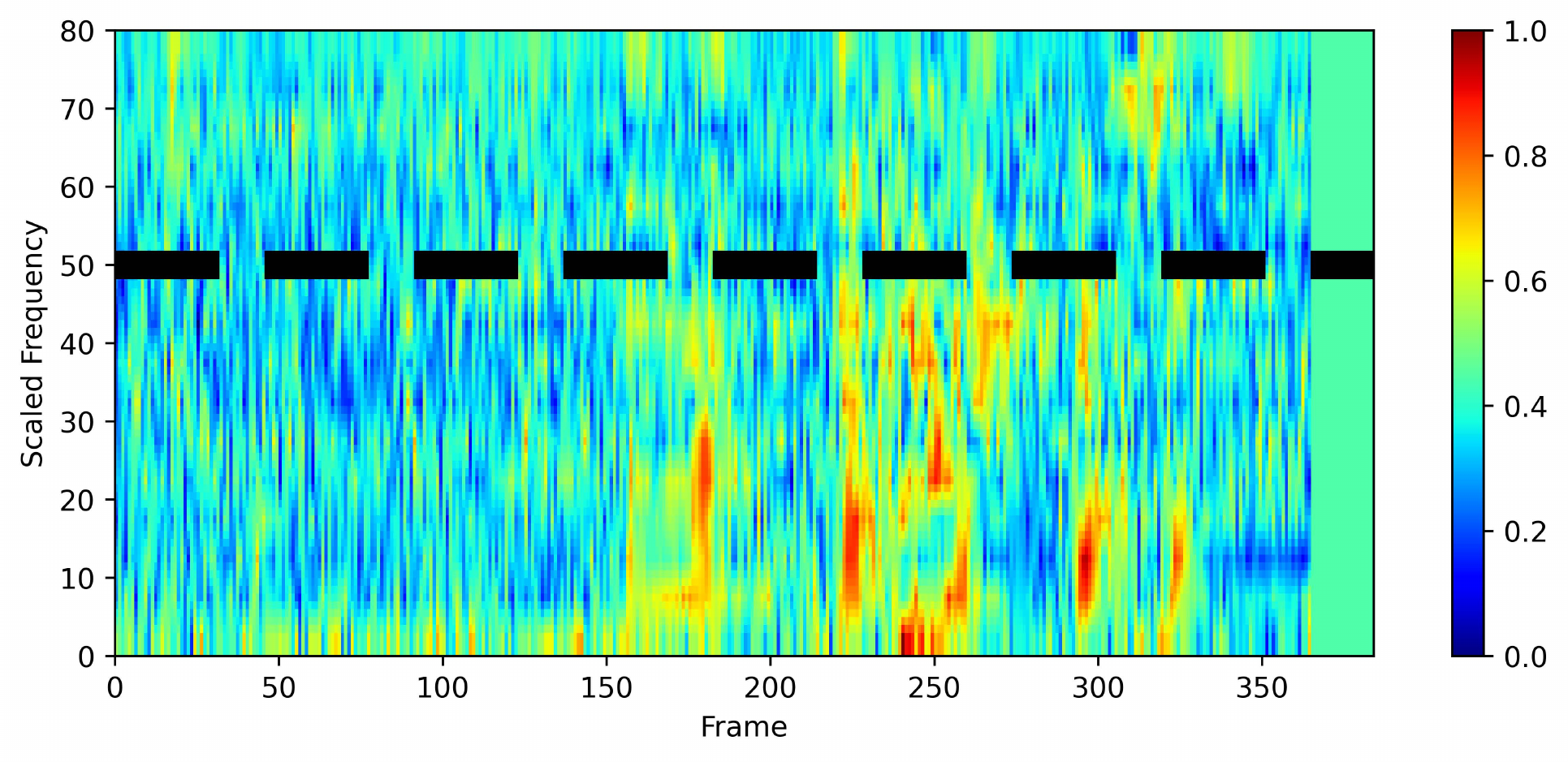}
\end{minipage}
\label{fig:vi-iem-sad-fat}
}
\subfigure[T-Attention of \emph{sad} on IEMOCAP]{
\begin{minipage}[b]{0.30\textwidth}
\includegraphics[width=\textwidth]{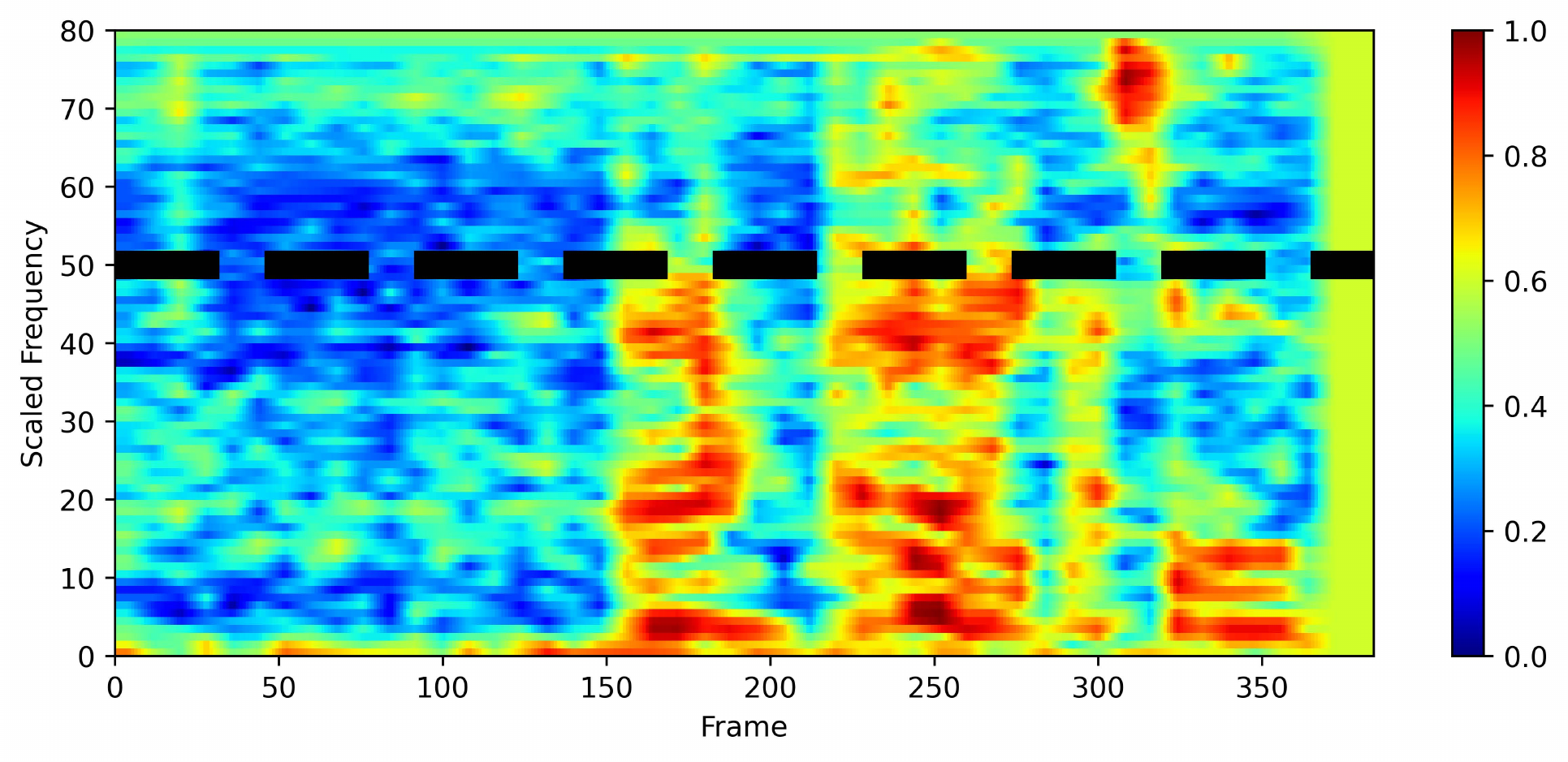}
\end{minipage}
\label{fig:vi-iem-sad-tat}
}

%
\subfigure[Spectrogram of \emph{cheerful} on ABC]{
\begin{minipage}[b]{0.30\textwidth}
\includegraphics[width=\textwidth]{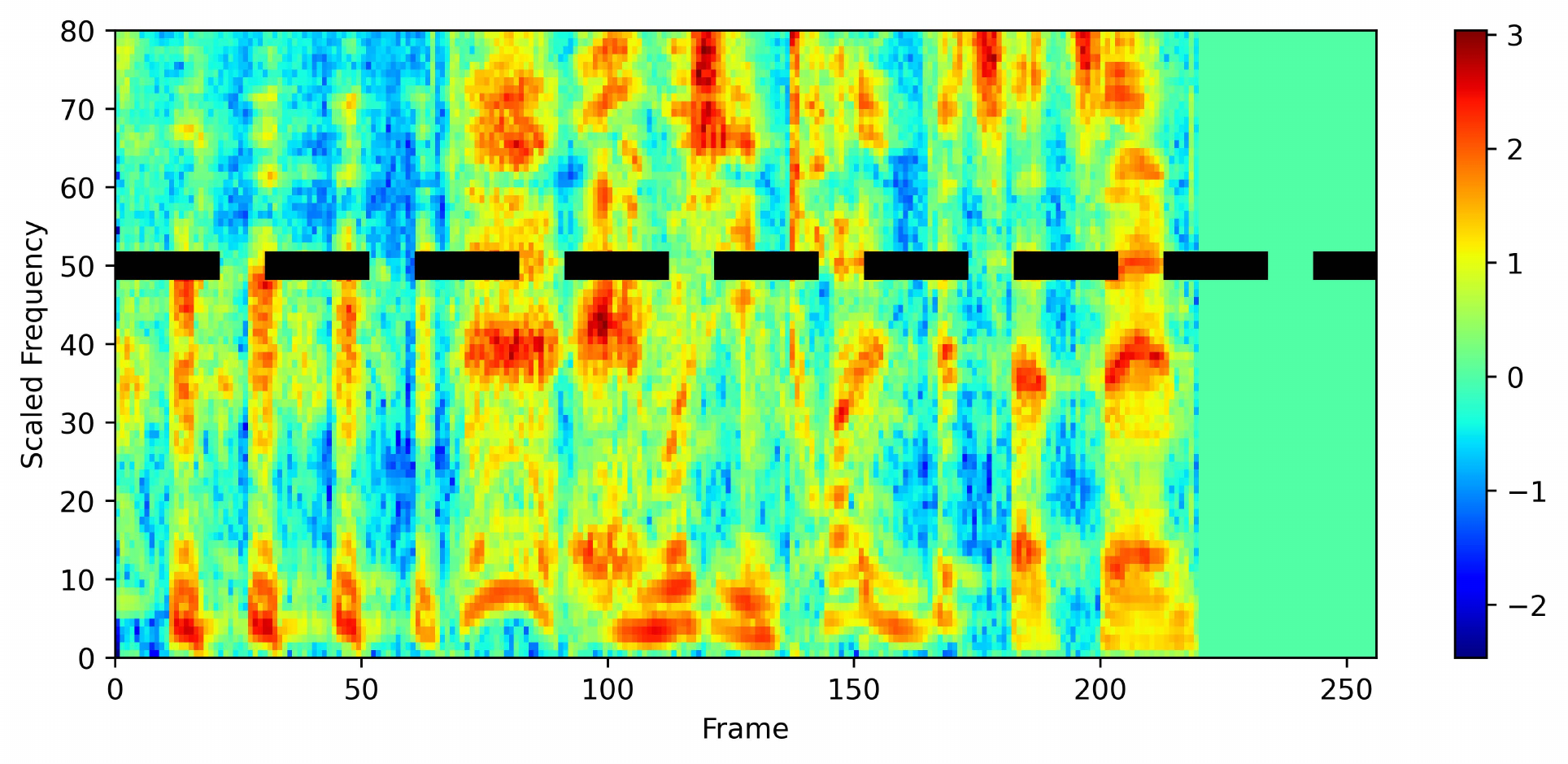}
\end{minipage}
\label{fig:vi-abc-che-spe}
}
\subfigure[F-Attention of \emph{cheerful} on ABC]{
\begin{minipage}[b]{0.30\textwidth}
\includegraphics[width=\textwidth]{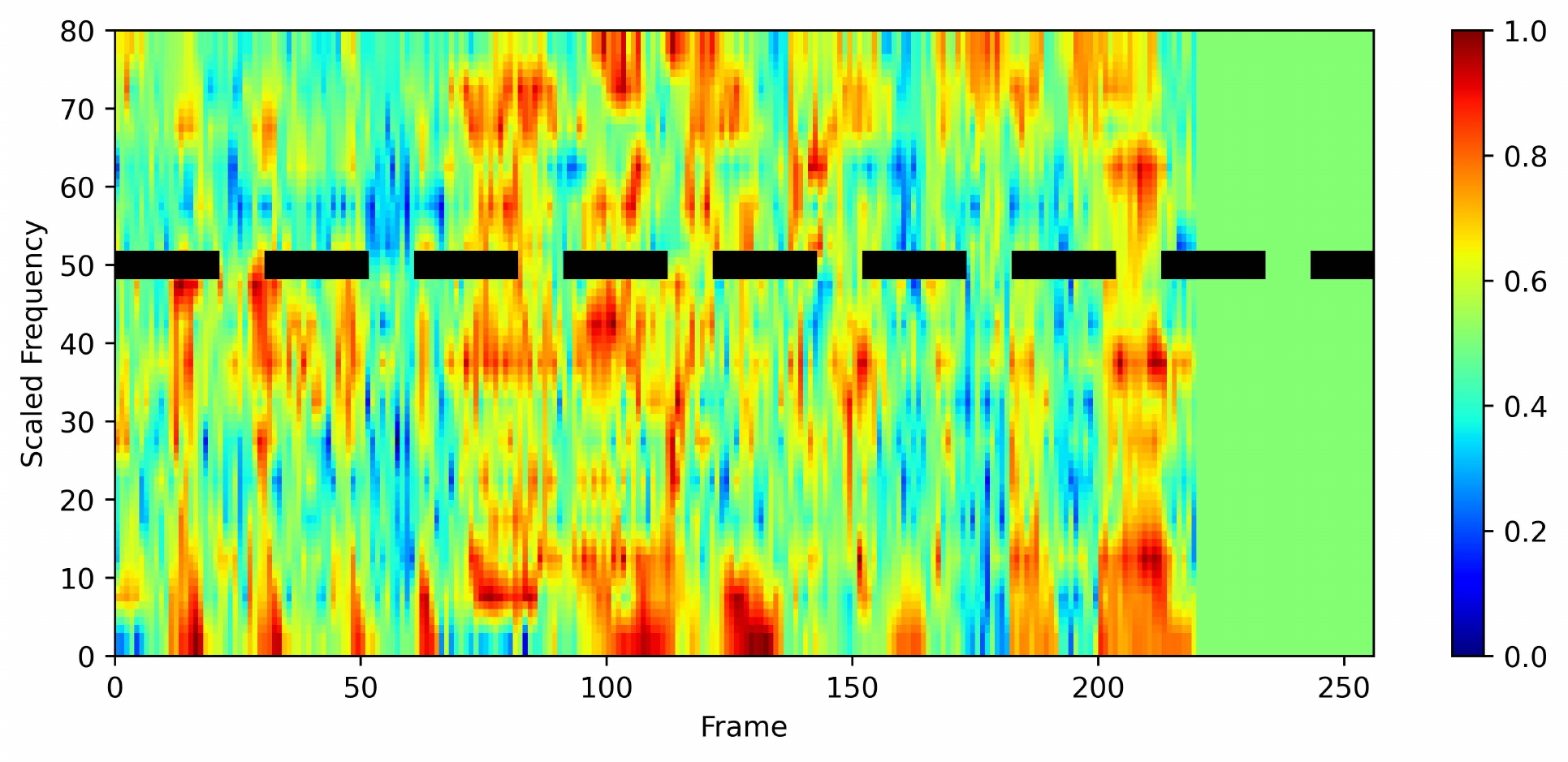}
\end{minipage}
\label{fig:vi-abc-che-fat}
}
\subfigure[T-Attention of \emph{cheerful} on ABC]{
\begin{minipage}[b]{0.30\textwidth}
\includegraphics[width=\textwidth]{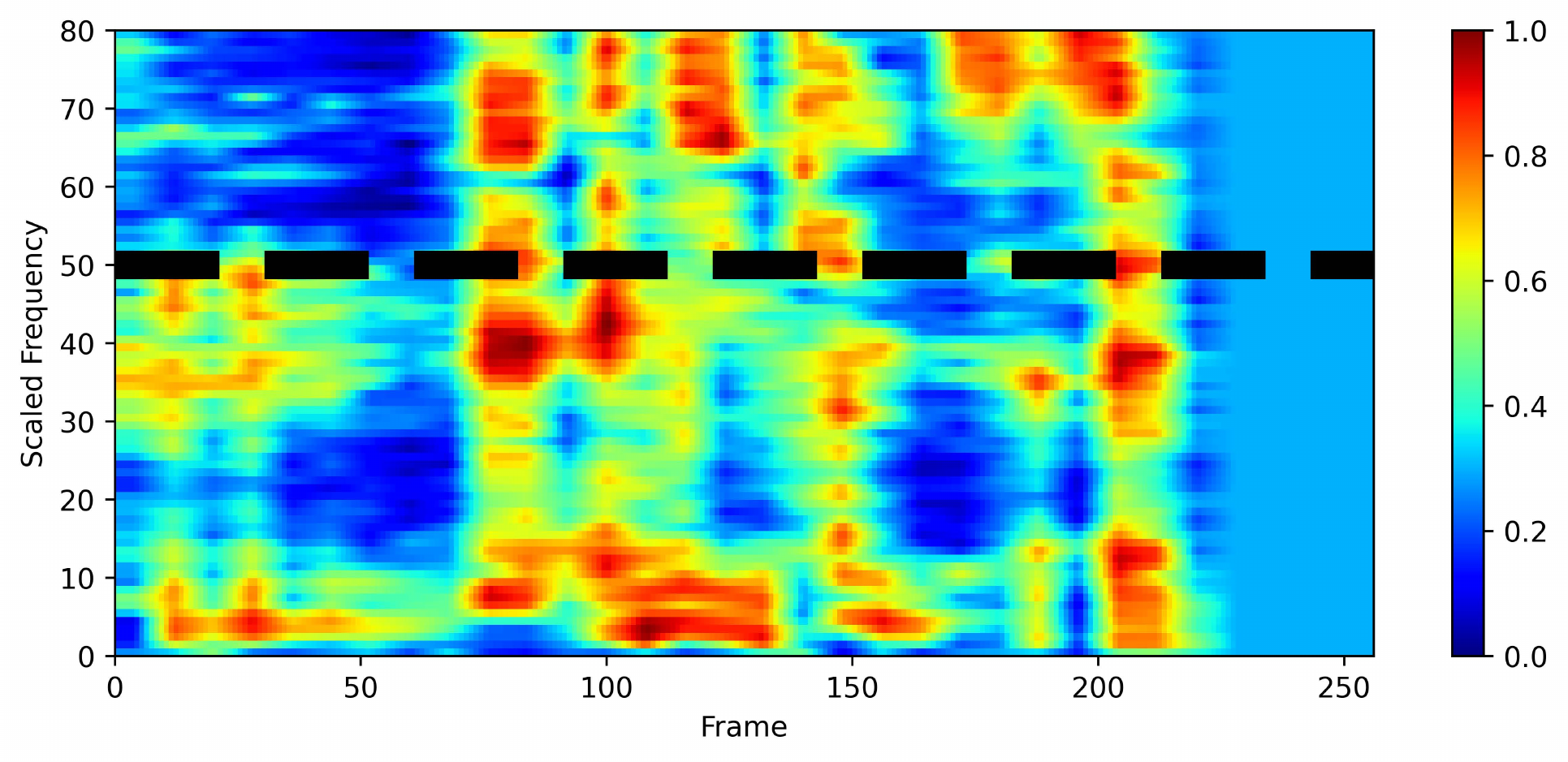}
\end{minipage}
\label{fig:vi-abc-che-tat}
}

\subfigure[Spectrogram of \emph{tired} on ABC]{
\begin{minipage}[b]{0.30\textwidth}
\includegraphics[width=\textwidth]{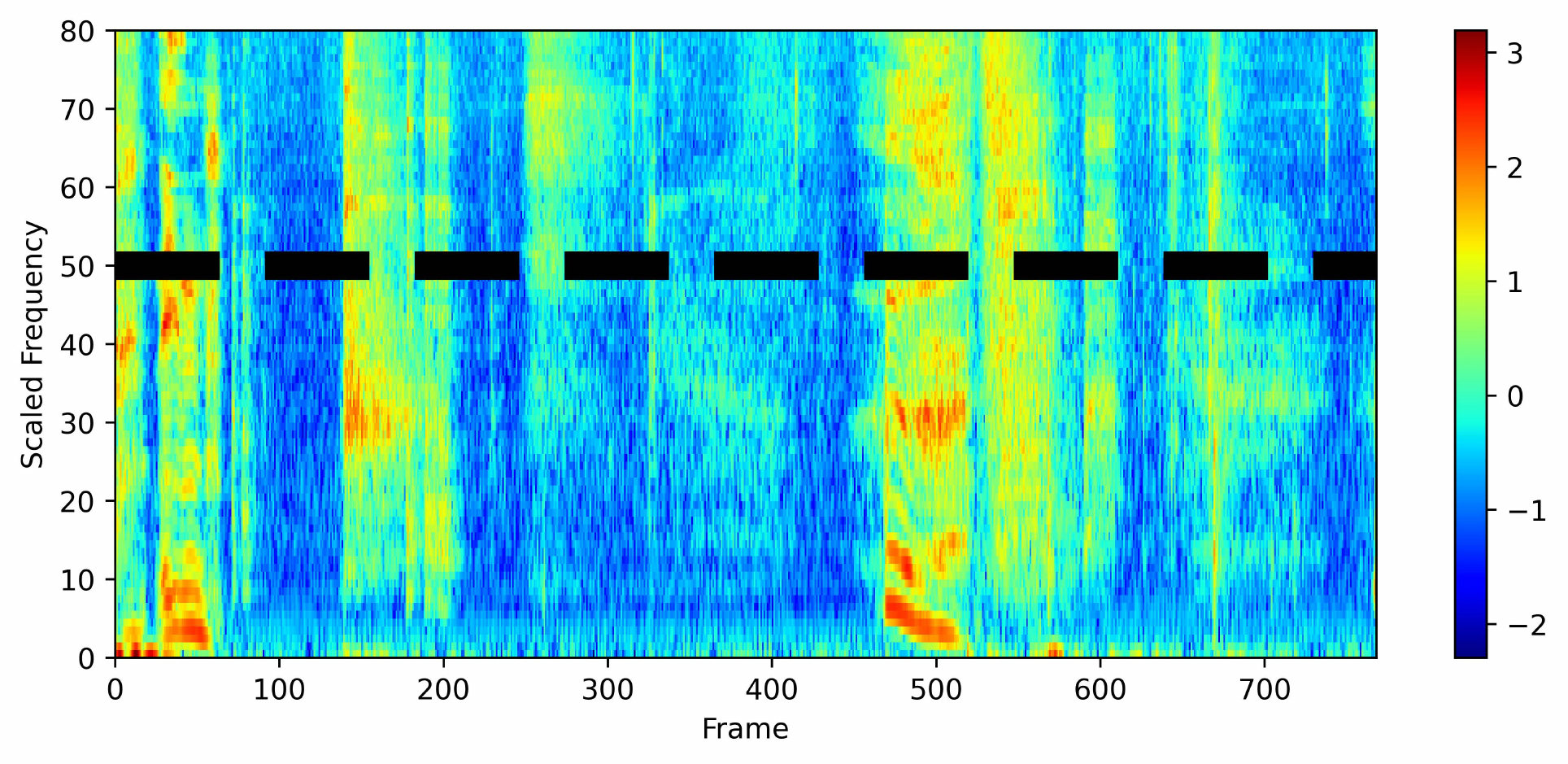}
\end{minipage}
\label{fig:vi-abc-tir-spe}
}
\subfigure[F-Attention of \emph{tired} on ABC]{
\begin{minipage}[b]{0.30\textwidth}
\includegraphics[width=\textwidth]{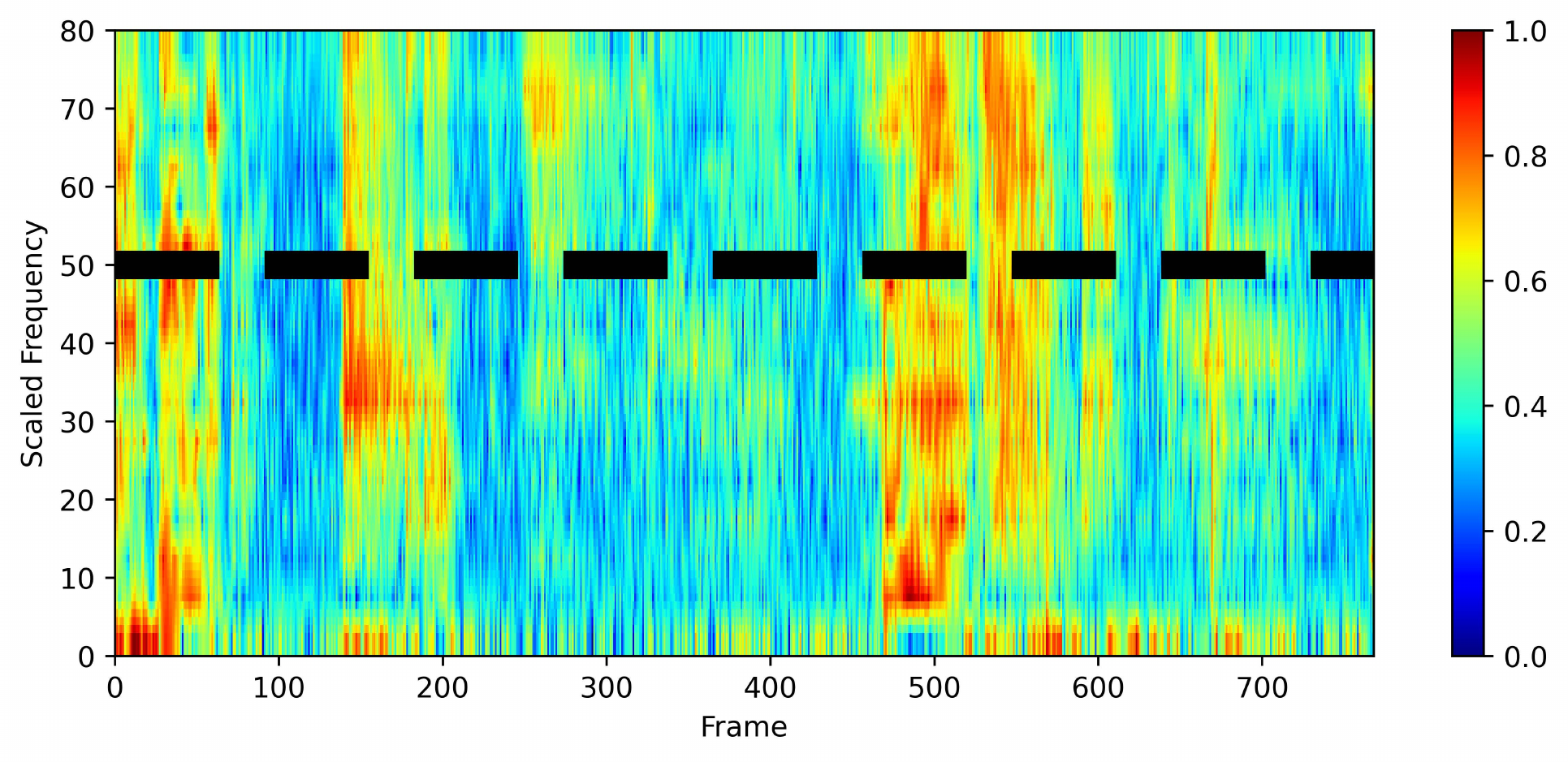}
\end{minipage}
\label{fig:vi-abc-tir-fat}
}
\subfigure[T-Attention of \emph{tired} on ABC]{
\begin{minipage}[b]{0.30\textwidth}
\includegraphics[width=\textwidth]{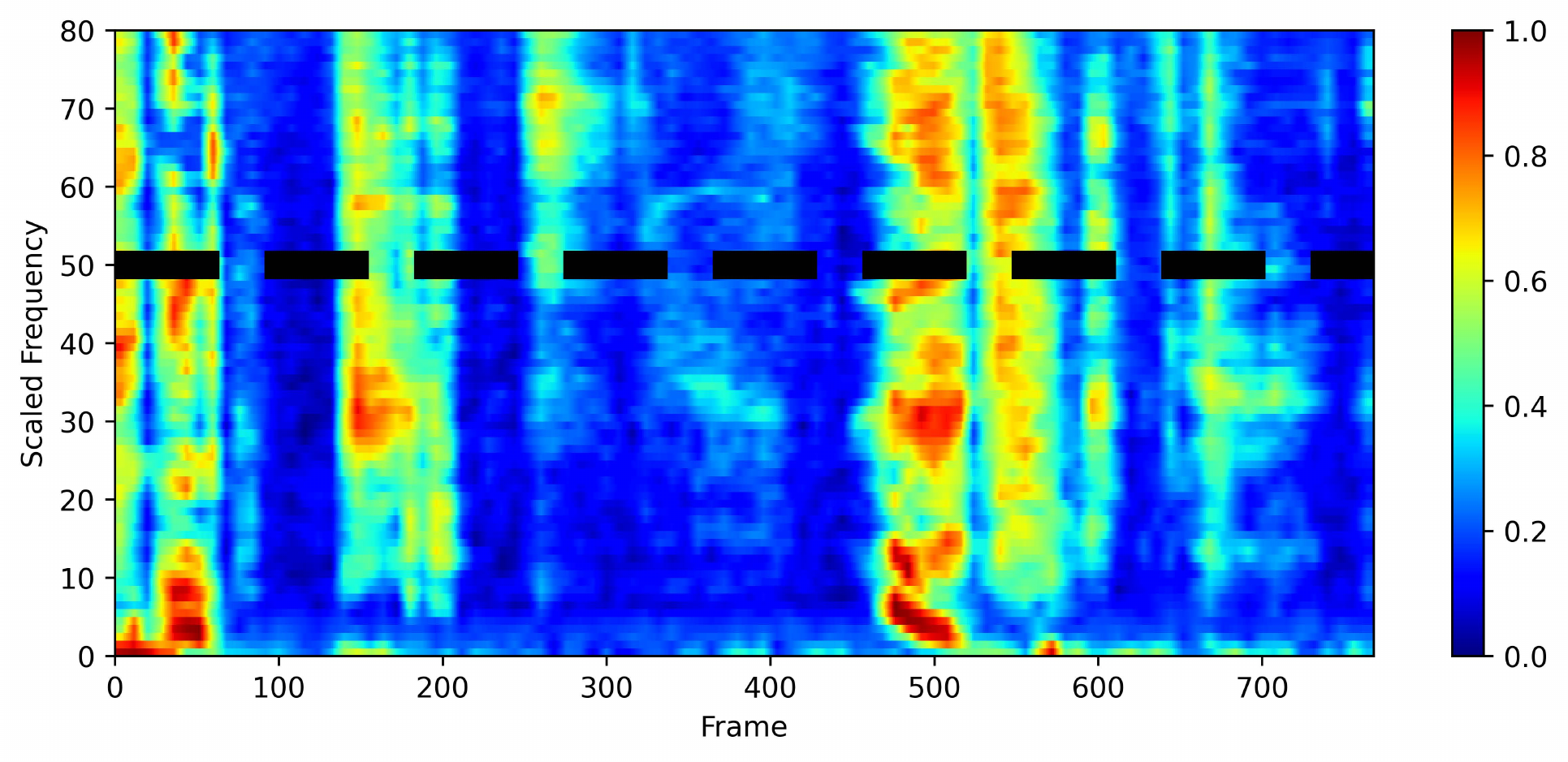}
\end{minipage}
\label{fig:vi-abc-tir-tat}
}

%
\subfigure[Spectrogram of \emph{happy} on CASIA]{
\begin{minipage}[b]{0.30\textwidth}
\includegraphics[width=\textwidth]{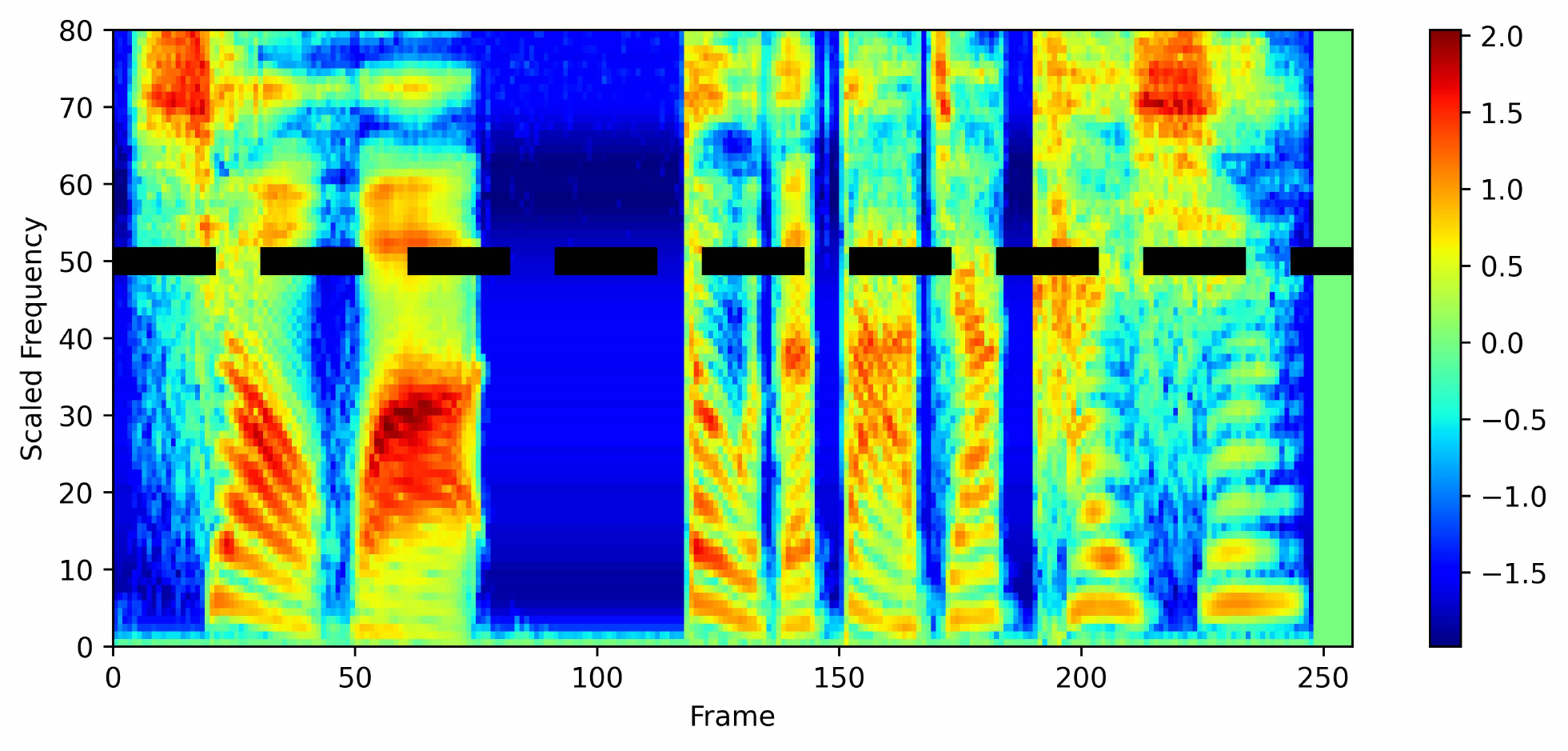}
\end{minipage}
\label{fig:vi-cas-hap-spe}
}
\subfigure[F-Attention of \emph{happy} on CASIA]{
\begin{minipage}[b]{0.30\textwidth}
\includegraphics[width=\textwidth]{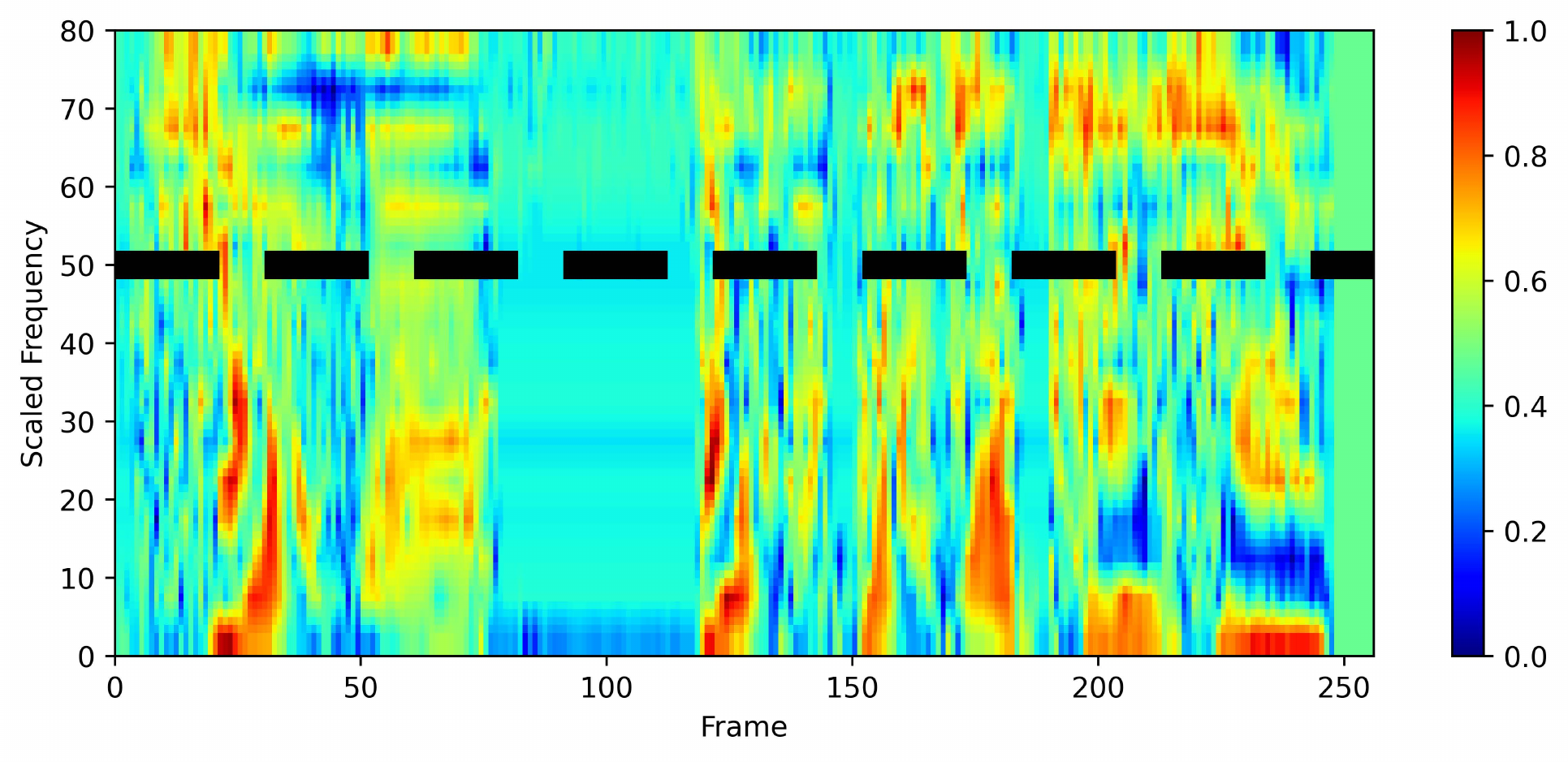}
\end{minipage}
\label{fig:vi-cas-hap-fat}
}
\subfigure[T-Attention of \emph{happy} on CASIA]{
\begin{minipage}[b]{0.30\textwidth}
\includegraphics[width=\textwidth]{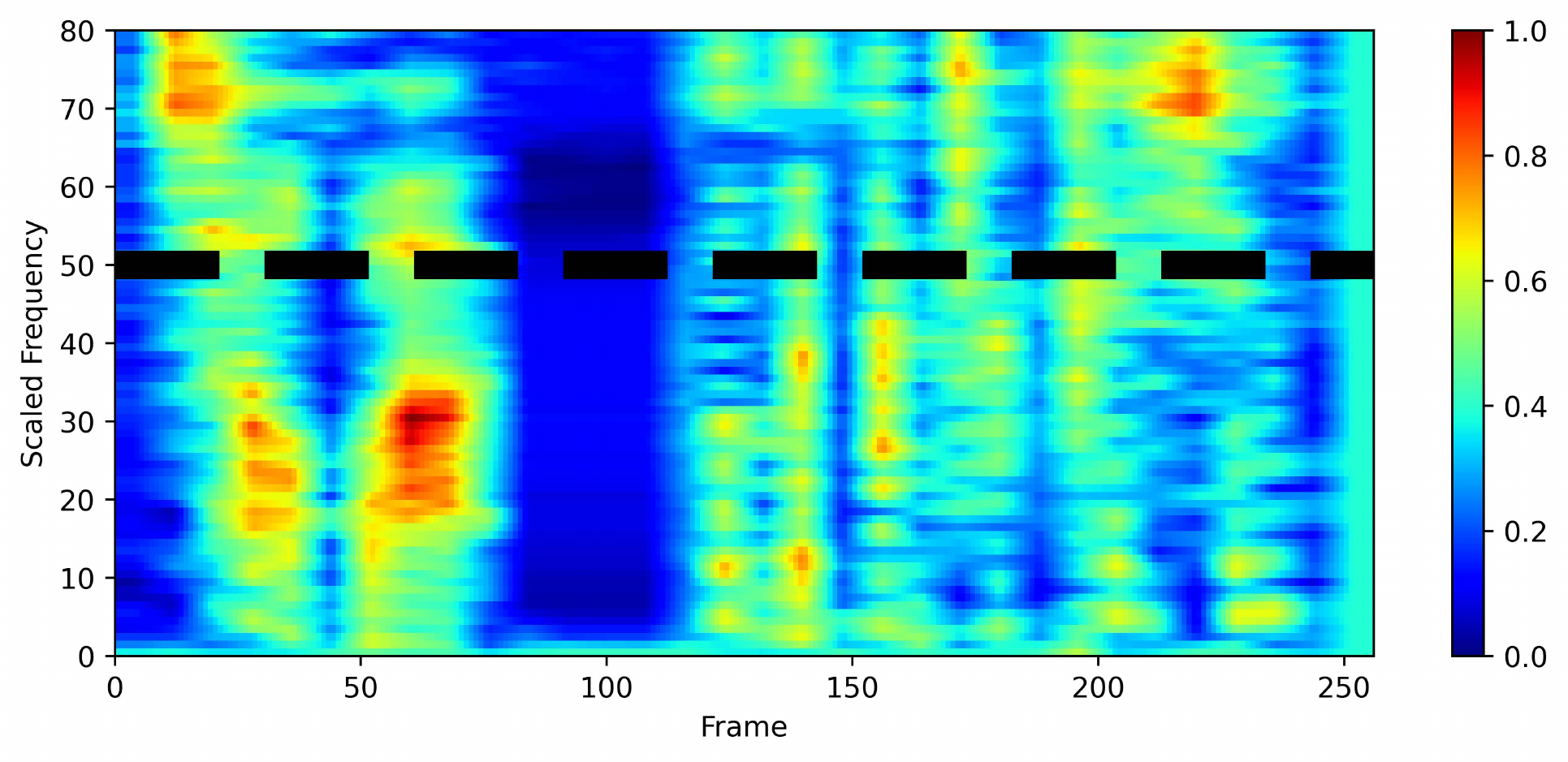}
\end{minipage}
\label{fig:vi-cas-hap-tat}
}

\subfigure[Spectrogram of \emph{sad} on CASIA]{
\begin{minipage}[b]{0.30\textwidth}
\includegraphics[width=\textwidth]{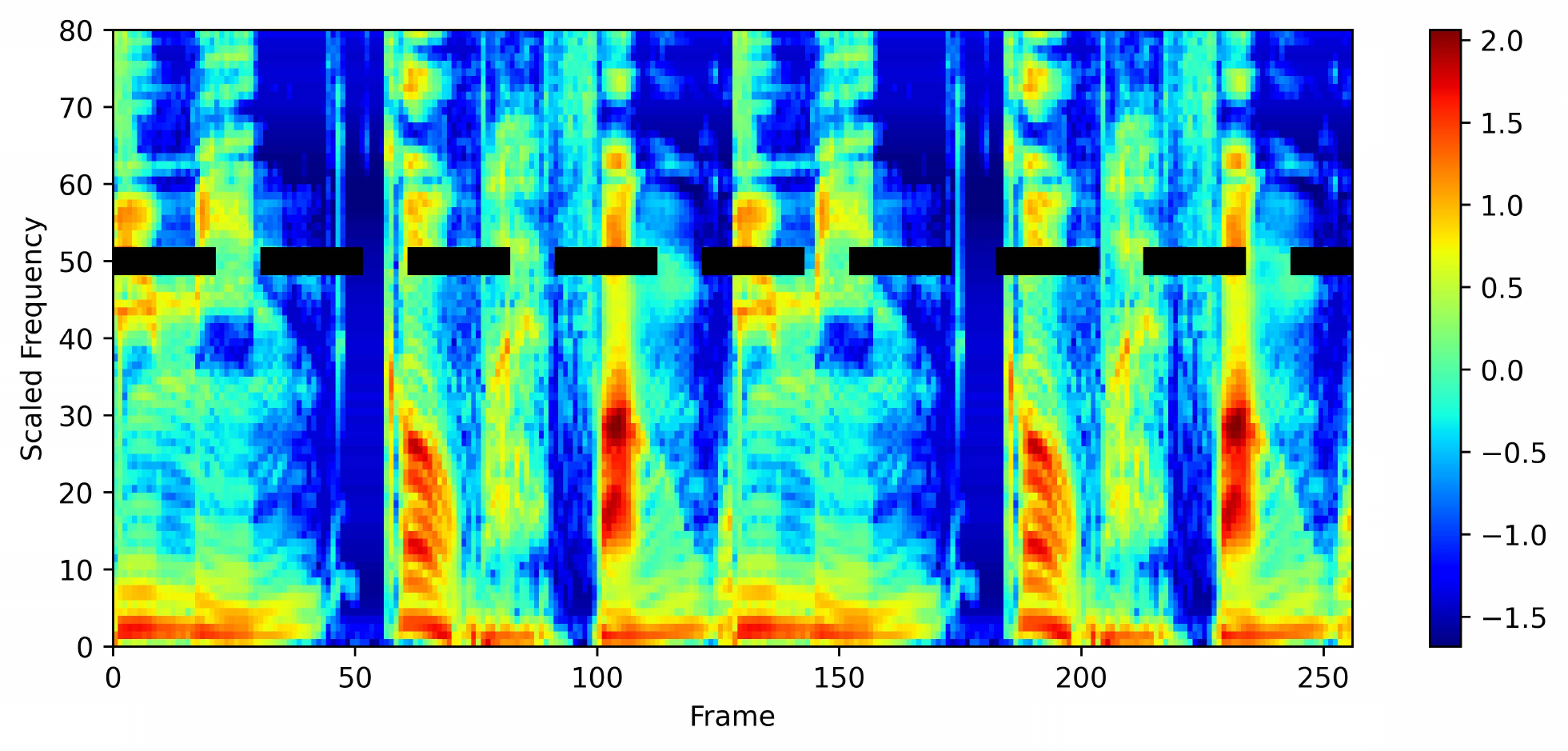}
\end{minipage}
\label{fig:vi-cas-sad-spe}
}
\subfigure[F-Attention of \emph{sad} on CASIA]{
\begin{minipage}[b]{0.30\textwidth}
\includegraphics[width=\textwidth]{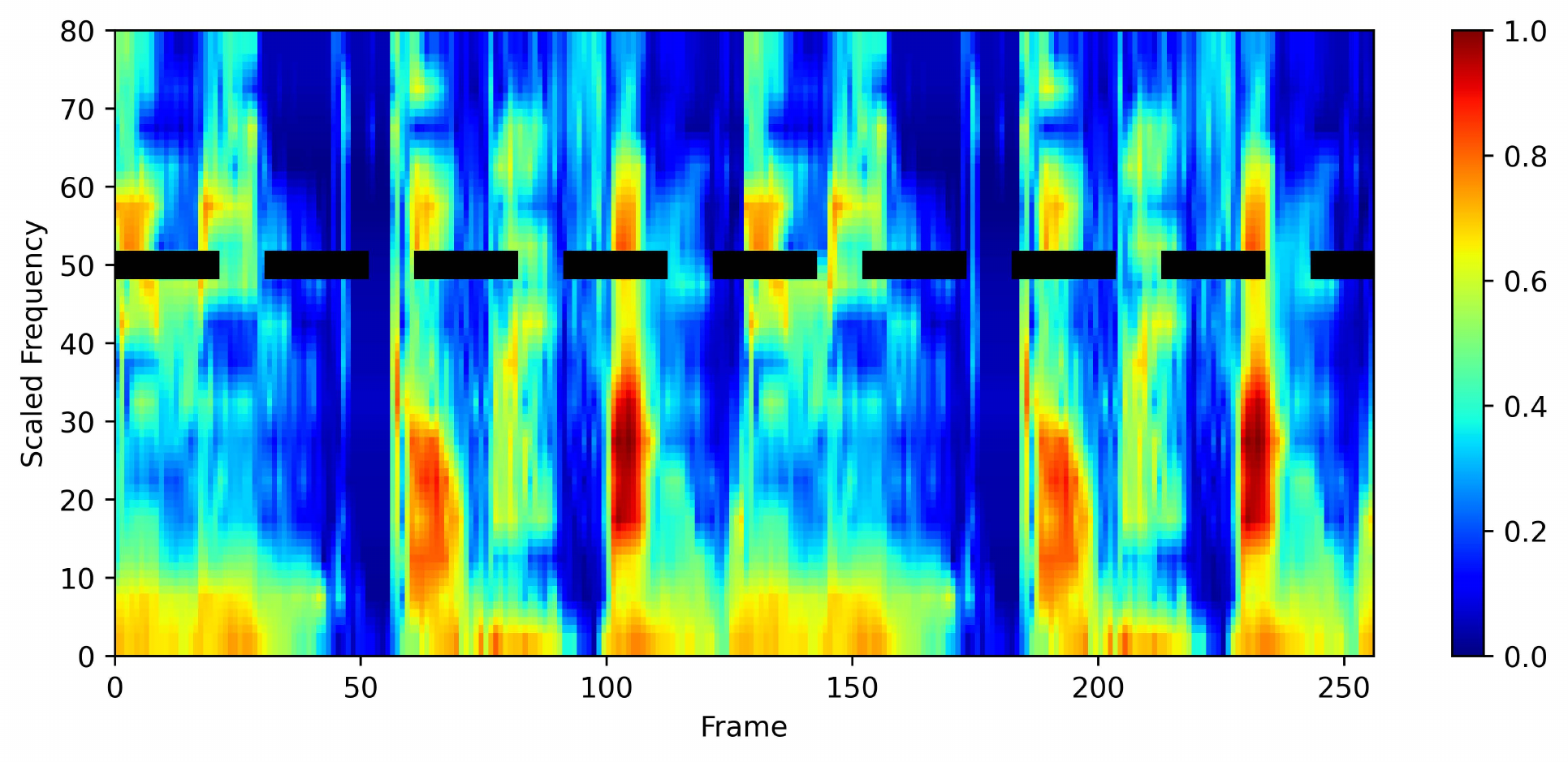}
\end{minipage}
\label{fig:vi-cas-sad-fat}
}
\subfigure[T-Attention of \emph{sad} on CASIA]{
\begin{minipage}[b]{0.30\textwidth}
\includegraphics[width=\textwidth]{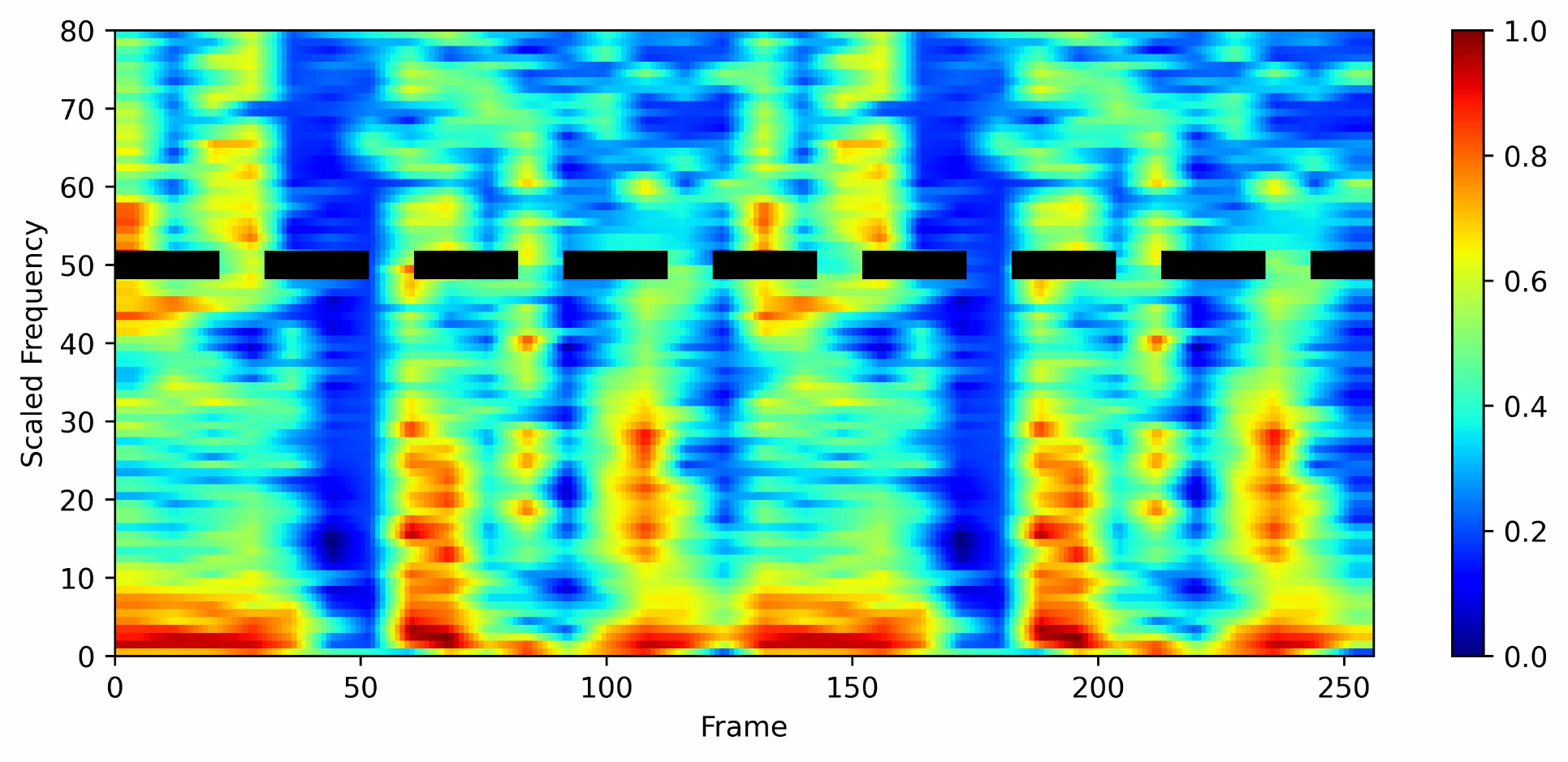}
\end{minipage}
\label{fig:vi-cas-sad-tat}
}

\caption{Visualization of log-Mel-spectrogram, T-Attention, and F-Attention under different emotions. (a)-(f) show the attentions under \emph{happy} and \emph{sad} on IEMOCAP. (g)-(l) show the attentions under \emph{cheerful} and \emph{tired} on ABC. (m)-(r) show the attentions under \emph{happy} and \emph{sad} on CASIA. In each figure, the upper part of the black dotted line '$\textbf{- -}$' means the high frequency energy activation region, while the lower part represents the middle and low frequency energy activation region.}
\label{fig:attention_visualization}
\end{figure*}

To further investigate the correlations of the frequency band activations on frequency domain and key frame regions on time domain, we visualize the F-Attention and the T-Attention on the log-Mel-spectrogram, as shown in Figure. \ref{fig:attention_visualization}. For this purpose, high-arousal and positive-valence emotions, e.\,g., \emph{happy}, \emph{cheerful}, and low-arousal and negative-valence emotions, e.\,g., \emph{sad}, \emph{tired}, are selected as comparison for attention visualization \cite{lang1995emotion}, \cite{kim2008emotion}. Among them, we choose \emph{happy} and \emph{sad} for a comparison on IEMOCAP and CASIA, whereas on \emph{cheerful} and \emph{tired}, which have close valence and arousal with \emph{happy} and \emph{sad}, are chosen as an alternative on ABC. Figure. \ref{fig:vi-iem-hap-spe}-\ref{fig:vi-iem-sad-tat} show the attentions under \emph{happy} and \emph{sad} on IEMOCAP, where Figure. \ref{fig:vi-iem-hap-spe}, \ref{fig:vi-iem-hap-fat}, and \ref{fig:vi-iem-hap-tat} respectively correspond to the visualizations of log-Mel-spectrogram, F-Attention, and T-Attention under \emph{happy}, while Figure. \ref{fig:vi-iem-sad-spe}, \ref{fig:vi-iem-sad-fat}, and \ref{fig:vi-iem-sad-tat}) are the visualizations of above three items under \emph{sad}. Similarly, the details of \ref{fig:vi-abc-che-spe}-\ref{fig:vi-cas-sad-tat} are given in Figure. \ref{fig:attention_visualization}. Notably, as shown on Figure. \ref{fig:attention_visualization}, the black dotted line '$\textbf{- -}$' on each figure represents the Mel-scaled frequency of 50, corresponding to the actual frequency band around 2\,700Hz. Therefore, the upper part of the black dotted line means the high frequency energy activation region, while the lower part represents the middle and low frequency energy activation region. \textcolor{black}{In addition, the attention visualizations reveal the specific activations under different emotions, where the colors of visualizations from red to blue represent corresponding activation intensity from large to small.}

\textcolor{black}{From the comparison of Figure. \ref{fig:vi-iem-hap-spe} and \ref{fig:vi-iem-hap-fat}, it is obvious to observe that there are frequent energy activations of \emph{happy} above 50 at the Mel-scaled frequency (corresponding to the frequency band around 2\,700Hz), which demonstrates that \emph{happy} has an increased in high-frequency energy. Whereas, as shown in Figure. \ref{fig:vi-iem-sad-spe} and \ref{fig:vi-iem-sad-fat}, we can clearly find that energy activations of \emph{sad} are infrequent above 50 at the Mel-scaled frequency, which indicates \emph{sad} has a decreased in high-frequency energy. Similar results also appear in ABC and CASIA, as shown in Figure. \ref{fig:vi-abc-che-spe} and \ref{fig:vi-abc-tir-spe}, Figure. \ref{fig:vi-cas-hap-spe} and \ref{fig:vi-cas-sad-spe}. These visualization results are consistent with the concepts reported in \cite{wu2011automatic}, \cite{cowie2001emotion}, \cite{kaiser1962communication}. Notably, the results in Figure. \ref{fig:vi-abc-tir-spe} and \ref{fig:vi-abc-tir-fat} also reveal that the F-Attention of \emph{tired} on ABC has frequent energy activations in the high-frequency bands from $10^{th}$ frame to $40^{th}$ frame and $125^{th}$ frame to $250^{th}$ frame. This is because the used speech sample has obvious sound of 'yawning' in these two frame intervals. On the contrary, the sample from $50^{th}$ frame to $80^{th}$ frame contains a conventional emotion expression of \emph{tired}, thus its F-Attention reveals infrequent energy activations above 50 Mel-scaled frequency, which is consist with the above demonstration. Moreover, the visualization of F-Attention confirms that our proposed method can focus on the frequency band activations related emotions, and then combine the correlations between these frequency bands into the frequency domain representation.}

\textcolor{black}{As for T-Attention, the Figure. \ref{fig:vi-iem-hap-tat} indicates that \emph{happy} on IEMOCAP has high activations in $50^{th}$ frame to $90^{th}$ frame and $110^{th}$ frame to $270^{th}$ frame. These frames correspond to the regions with rich speech contents shown in Figure. \ref{fig:vi-iem-hap-spe}, which are all key regions in frames that highly contribute to speech emotion representation. From the visualization of T-Attention on other emotions or databases, i.\,e., Figure. \ref{fig:vi-iem-sad-tat}, \ref{fig:vi-abc-che-tat}, \ref{fig:vi-abc-tir-tat}, \ref{fig:vi-cas-hap-tat}, and \ref{fig:vi-cas-sad-tat}, it is clear to discover similar results with Figure. \ref{fig:vi-iem-hap-tat}. Also, our proposed T-Attention has few activations on all silent frames, e.\,g., $0^{th}$ frame to $150^{th}$ frame and $360^{th}$ frame to $384^{th}$ frame in Figure. \ref{fig:vi-iem-sad-spe}, $220^{th}$ frame to $256^{th}$ frame in Figure. \ref{fig:vi-abc-che-spe}, $80^{th}$ frame to $120^{th}$ frame and $210^{th}$ frame to $480^{th}$ frame in Figure. \ref{fig:vi-abc-tir-spe}, $70^{th}$ frame to $120^{th}$ frame in Figure. \ref{fig:vi-cas-hap-spe}, $40^{th}$ frame to $60^{th}$ frame in Figure. \ref{fig:vi-cas-sad-spe}. These salient results further emphasize that our proposed T-Attention effectively focuses on the key regions related to emotions in time frames.}

\section{Conclusion}

In the paper, we propose an attentive time-frequency neural network (ATFNN) to extract the discriminative presentation for the SER. We first utilize a time-frequency neural network (TFNN), integrating a domain encoder and a time domain encoder into the time-frequency joint learning strategy, to model the correlations within frequency bands and within time frames. Then, a frequency attention network and a time attention network are also embedded into the TFNN to capture the specific frequency bands and time frames related to emotions. Experimental results on the two public databases, i.\,g., IEMOCAP and ABC, prove that our ATFNN outperforms the state-of-the-art methods. Furthermore, the visualization of attentions reveals that the ATFNN can focus on the regions related to emotions of the frequency domain and the time domain. Moreover, precise correspondence between different emotions and frequency bands should be deeply explored in future research to improve the robustness and discriminability of speech emotion features.


%

%
%
%
%

\ifCLASSOPTIONcaptionsoff
  \newpage
\fi

\bibliographystyle{IEEEtran}
\bibliography{reference}

\end{document}